\documentclass[11pt]{article}
\usepackage[labelfont=bf]{caption} % Bold figure number
\usepackage[font={small}]{caption}   %% FIGURE CAPTION FONT MODIFICATION

\usepackage{graphicx}  % needed for figures
\usepackage{dcolumn}   % needed for some tables
\usepackage{bm}        % for math
\usepackage{amssymb}   % for math

\usepackage{upgreek}	   % upright Greek

\usepackage{amsmath}   % for math align environment
\usepackage{paralist}  % for compactitem and compactenum

\usepackage[numbers, sort&compress]{natbib}

\usepackage[labelfont=bf]{caption} % Bold figure number
\usepackage[strict]{changepage} % Adjust figure and table width
\usepackage{textgreek} % Non-italic Greek letters

\usepackage{lineno}

\usepackage[usenames,dvipsnames]{color}    % ADDED FOR COLOR NAMES
\usepackage{soul}   % SOUL Added for underlines
\setstcolor{red} % strikethrough color
\setul{}{1.5pt}  % line thickness

\DeclareRobustCommand{\rchi}{{\mathpalette\irchi\relax}}
\newcommand{\irchi}[2]{\raisebox{\depth}{$#1\chi$}} % inner command, used by \rchi

\newcommand*\patchAmsMathEnvironmentForLineno[1]{%
  \expandafter\let\csname old#1\expandafter\endcsname\csname #1\endcsname
  \expandafter\let\csname oldend#1\expandafter\endcsname\csname end#1\endcsname
  \renewenvironment{#1}%
     {\linenomath\csname old#1\endcsname}%
     {\csname oldend#1\endcsname\endlinenomath}}% 
\newcommand*\patchBothAmsMathEnvironmentsForLineno[1]{%
  \patchAmsMathEnvironmentForLineno{#1}%
  \patchAmsMathEnvironmentForLineno{#1*}}%
\AtBeginDocument{%
\patchBothAmsMathEnvironmentsForLineno{equation}%
\patchBothAmsMathEnvironmentsForLineno{align}%
\patchBothAmsMathEnvironmentsForLineno{flalign}%
\patchBothAmsMathEnvironmentsForLineno{alignat}%
\patchBothAmsMathEnvironmentsForLineno{gather}%
\patchBothAmsMathEnvironmentsForLineno{multline}%
}

\begin{document}

\title{\textbf{Dynamic suppression of Rayleigh light scattering in dielectric resonators}}
	
\author{Seunghwi Kim$^1$, Jacob M. Taylor$^{2,3}$, and Gaurav Bahl$^{1\ast}$\\
	\\
	\footnotesize{$^1$ Mechanical Science and Engineering, University of Illinois at Urbana-Champaign,}\\
	\footnotesize{Urbana, Illinois 61801, USA}\\
	\footnotesize{$^2$ Joint Quantum Institute, University of Maryland, }\\
	\footnotesize{College Park, Maryland 20742, USA}\\
	\footnotesize{$^3$ Joint Center for Quantum Information and Computer Science,} \\
	\footnotesize{National Institute of Standards and Technology, Gaithersburg, Maryland 20899, USA}\\
	\footnotesize{$^\ast$ To whom correspondence should be addressed; bahl@illinois.edu} \\
}
	
\date{}
	
	\vspace*{-2cm}
	{\let\newpage\relax\maketitle}

%%%%%%%%%%%%%%%%%%%%%%%%%%
%% Abstract here
%%%%%%%%%%%%%%%%%%%%%%%%%%

\textbf{\small{
The ultimate limits of performance for any classical optical system are set by sub-wavelength fluctuations within the host material, that may be frozen-in or even dynamically induced. 
The most common manifestation of such sub-wavelength disorder is Rayleigh light scattering, which is observed in nearly all wave-guiding technologies today and can lead to both irreversible radiative losses 	
as well as undesirable intermodal coupling 
\cite{Marcuse:1969, Pinnow:1973, Weiss:1995, Gorodetsky:2000}. 	
While it has been shown that backscattering from disorder can be suppressed by breaking time-reversal symmetry \cite{Golubentsev:1984,John:1988} in magneto-optic and topological insulator materials \cite{Lenke:2000, Buttiker:1988, Haldane:2008, ZhengWang:2009}, common optical dielectrics possess neither of these properties.
Here we demonstrate an optomechanical approach for dynamically suppressing Rayleigh backscattering within dielectric resonators.
We achieve this by locally breaking time-reversal symmetry in a silica resonator through a Brillouin scattering interaction that is available in all materials.
Near-complete suppression of Rayleigh backscattering is experimentally confirmed through three independent measurements -- the reduction of the back-reflections caused by scatterers, the elimination of a commonly seen normal-mode splitting effect, and by measurement of the reduction in intrinsic optical loss.
More broadly, our results provide new evidence that it is possible to dynamically suppress Rayleigh backscattering within any optical dielectric medium, for achieving robust light propagation in nanophotonic devices in spite of the presence of scatterers or defects.
}}

\newpage

%%%%%%%%%%%%%%%%%%%%%%%%%%
%% Main paper begins
%%%%%%%%%%%%%%%%%%%%%%%%%%

%%%%%%%%%%%%%%%%%%%%%%%%%%
%% Introduction
%%%%%%%%%%%%%%%%%%%%%%%%%%

Rayleigh scattering is routinely encountered in nanostructured photonic devices as it limits microresonator quality (Q) factors 
	\cite{Weiss:1995, Gorodetsky:2000, Mazzei:2007}, 
affects the stability of frequency combs \cite{Griffith:2015, Suh:2016}, causes Anderson localization \cite{Schwartz:2007}, and limits the performance of metasurfaces \cite{Nagpal:2009}.
It can be induced by inhomogeneities in the form of internal stresses, point defects, density variations, dislocations, and even surface roughness, which are unavoidable due to manufacturing limitations but may also occur thermodynamically. 
In particular, back-reflections arising from Rayleigh scatterers in nanostructured devices create prominent reflections in silicon photonics 
\cite{Morichetti:2010} 
and a well-known mode splitting or `doublet' phenomenon in resonators \cite{Weiss:1995, Gorodetsky:2000, Mazzei:2007, Kippenberg:2009}, both of which impose severe technological constraints.

An elegant proposal to counteract disorder-induced backscattering of electromagnetic waves is to break the time-reversal symmetry (TRS) of the medium \cite{Golubentsev:1984,John:1988} -- so that modes available for opposite, i.e., time-reversed, propagation are simply not symmetric in energy-momentum space. In other words, backscattering can be suppressed by establishing a large contrast in the optical density of states for propagation in the opposing directions. This effect has been experimentally confirmed in Faraday rotator (magneto-optic) materials biased with large magnetic fields \cite{Lenke:2000} but cannot be extended to common dielectrics. 
A similar effect in which broken TRS suppresses electron backscattering is also seen in the chiral edge currents of two-dimensional electron systems exhibiting the quantum Hall effect (QHE) \cite{Halperin:1982, Buttiker:1988}. 
More recently, there has been a flurry of activity on backscattering suppression via TRS-breaking in photonic topological insulator metamaterials after the analogy to the QHE was established 
\cite{Haldane:2008},  
with successful demonstration in a magneto-optic photonic crystal \cite{ZhengWang:2009} and through Floquet pumping 
\cite{Rechtsman:2013}..
Unfortunately, since common photonic materials do not have magneto-optical activity and are topologically trivial insulators in their band gaps, how these lessons may be mapped to any monolithic dielectric waveguide remains an open question.

In this work, we demonstrate a simple optomechanical approach by which we can dynamically suppress electromagnetic backscattering in any dielectric. The approach relies on an induced transparency process supported by Brillouin light scattering, which is a high gain optical nonlinearity available in all phases of matter, and has been established as a highly effective tool for breaking TRS in dielectric waveguides and resonators \cite{Kang:2011, Kim:2015, Kim:2017}. 
Using this technique, we experimentally demonstrate near-complete suppression of Rayleigh backscattering within monolithic silica microresonators, with dynamic control provided by an external optical pump. The effect is confirmed both through elimination of back-reflected light as well as elimination of the normal mode splitting between cw and ccw modes of the symmetric resonators. Our experiments exhibit a restoration to the intrinsic material loss rate of an optical resonator in spite of the presence of scattering defects.

\begin{figure}[tp!]
	\begin{adjustwidth}{-1in}{-1in}
	\makebox[\textwidth][c]{\includegraphics[width=1.2\textwidth]{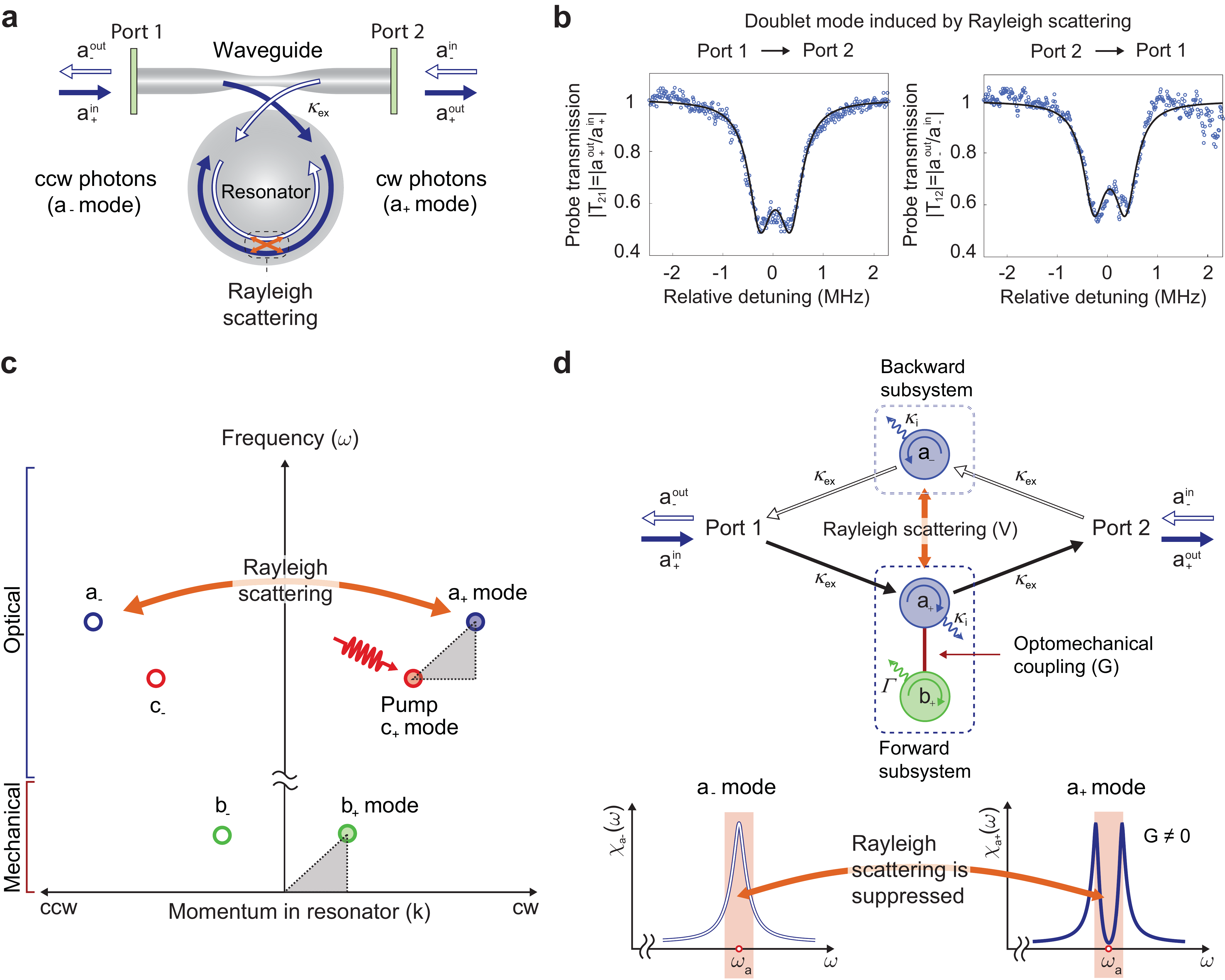}}
	\centering
	\caption{
	\textbf{Rayleigh backscattering in a whispering gallery resonator (WGR) and concept for optomechanical suppression.} 
	\textbf{(a)} Optical WGRs support degenerate modes ($a_\pm$) that are time-reversed partners (cw/ccw) and can be individually accessed via directional probing. However, Rayleigh backscattering from disorder intrinsic to the WGR can couple these modes, leading to loss of their distinguisable directionality.
	\textbf{(b)} Experimentally, this can result in normal mode splitting or `doublet' (measured here in a silica WGR) when the disorder induced backscattering rate is comparable to the intrinsic optical loss rate. 
	Such doublets are routinely observed in high-Q resonator systems and impose a technological constraint.
	\textbf{(c)} We can suppress Rayleigh backscattering by breaking time-reversal symmetry within the bandwidth of the $a_\pm$ optical modes. This is achieved through a Brillouin optomechanical induced transparency process \cite{Kim:2015, Kim:isolator}, in which a high-coherence cw mechanical mode $b_+$ is coupled to the cw $a_+$ mode by a cw directional pump $c_+$. The interaction is subject to the phase matching constraint illustrated by the grey triangle. The momentum matching requirement implies that the cw pump does not directly induce any effect for the ccw optical mode $a_-$.
	\textbf{(d)} Toy model for the WGR and waveguide system in which we distinguish the two directional subsystems and indicate both Rayleigh ($V$) and optomechanical ($G$) couplings. All variables are defined in the main text. The directional Brillouin optomechanical coupling significantly reduces the susceptibility of the $a_+$ mode only and `open-circuits' the backscattering channel, thereby suppressing Rayleigh scattering.
	}
	\label{fig:1}
	\end{adjustwidth}
\end{figure}

%%%%%%%%%%%%%%%%%%%%%%%%%%
%% Introduction on our experimental system
%%%%%%%%%%%%%%%%%%%%%%%%%%

\vspace{12pt}

The system under consideration (Fig.~\ref{fig:1}a) is a symmetric whispering gallery resonator (WGR) that supports two degenerate optical modes $a_\pm$ associated with clockwise (+ or cw) and counter-clockwise (- or ccw) photon propagation.
Such degenerate modes, which are time-reversed partners, can be coupled through backscattering from inhomogeneities or defects, resulting in a doublet mode if the coupling rate is comparable to the optical loss rate \cite{Weiss:1995, Gorodetsky:2000, Mazzei:2007}. For lower backscattering rates the mode only appears slightly broadened from its intrinsic linewidth.
Fig.~\ref{fig:1}b exhibits this backscattering induced mode split measured in a silica microsphere WGR. The measurement is performed via evanescent probing through a tapered fiber waveguide (Fig.~\ref{fig:1}a) such that the optical resonances appear as a dip in transmission. Optical probing in both directions through the waveguide indicates that the $a_\pm$ modes are hybridized due to Rayleigh backscattering \cite{Weiss:1995, Gorodetsky:2000, Mazzei:2007} and have lost their distinguishable directionality. It is this detrimental mode splitting and broadening effect that we wish to mitigate.

In order to experimentally shut down the effect of the backscattering channel between $a_+$ and $a_-$ 
we break time-reversal symmetry within the bandwidth of these modes. Specifically we employ Brillouin scattering induced transparency (BSIT) \cite{Kim:2015}, 
which is a non-reciprocal process that allows us to only modify the susceptibility $\rchi_{a+}(\omega)$ of the $a_+$ mode while leaving the $a_-$ mode nominally unaffected (Fig.~\ref{fig:1}d-bottom).
As with other optomechanically-induced transparencies 
	\cite{Weis_OMIT, Safavi-Naeini_OMIT}, 
the BSIT arises due to coherent coupling 
	\cite{Groblacher:2009} % here
between an optical mode with long-lived mechanical state, that is enabled through radiation forces and photoelastic scattering. When the coupling rate between light and the mechanics is sufficiently large, the optical mode exhibits normal mode splitting -- hybridization of the mechanical and optical modes -- which inhibits on-resonance absorption from the waveguide.
Production of BSIT requires a Stokes-detuned pump optical field (on a different optical mode $c_+$) that co-propagates with the mode of interest $a_+$, and a high-Q mechanical whispering gallery mode $b_+$ within the WGR. These three modes must together be subject to the Brillouin phase matching condition on both frequency $\omega_a = \omega_c + \omega_b$ and momentum $k_a = k_c + k_b$ as illustrated in Fig.~\ref{fig:1}c.
In our experiment we use $c_+$ pumping only, 
although a $c_-$ pump could also be invoked to independently control the susceptibility for the $a_-$ mode~\cite{Kim:isolator}.
It is the unique, momentum-selective feature of BSIT that allows us to break TRS for light propagation within the resonator.

\vspace{12pt}

The model Hamiltonian for this system includes both Rayleigh backscattering and the optomechanical interaction as follows:
\begin{align}
	H_{\text{int}} = \hbar(G a_+^{\dagger} b_+ + G^* a_+ b_+^{\dagger}) + \hbar V (a_+^{\dagger} a_-  + a_-^{\dagger} a_+) ~.
\end{align}
Here, $V$ is the Rayleigh backscattering induced coupling rate between the $a_\pm$ modes, while $G = g_o \sqrt{n_{c+}}$ is the pump-enhanced clockwise-only optomechanical coupling rate between the $a_+$ optical mode and the $b_+$ mechanical mode. $g_o$ represents the single photon optomechanical coupling rate, and $n_{c+}$ represents the average number of intracavity photons in the $c_+$ mode.  Since no ccw pump is applied to the system, an optomechanical interaction between $a_-$ and $b_-$ need not be considered.  
A detailed analysis presented in Supplement \S S1 additionally incorporates the effects of disorder-induced backscattering within the pump modes $c_\pm$, which can be distinct from the Rayleigh coupling between $a_\pm$ due to differences of modeshape and polarization. However as we show in Supplement \S S1.6 - \S S1.7 there is no evidence for this additional scattering effect within the pump modes, in the experiments that we discuss in this paper.

Fig.~\ref{fig:1}d presents a toy model of the system, in which we explicitly distinguish between forward and backward subsystems.
The degenerate optical modes $a_\pm$ are modeled with intrinsic loss rate $\kappa_i$, which includes all absorption and scattering mechanisms that leak light out of the mode, but excludes the influence of the backscattering channel $V$. These modes $a_\pm$ couple to the waveguide with an extrinsic coupling rate defined by $\kappa_{\text{ex}}$ that is symmetric in both forward and backward directions.
For light propagating in the waveguide from \mbox{Port 1} $\rightarrow$ Port 2 (forward direction), the interaction with the resonator occurs through the forward subsystem described as the optomechanically coupled $a_+$ optical and $b_+$ mechanical modes. 
Conversely, for light propagating from Port 2 $\rightarrow$ Port 1 (backward direction), light primarily interacts with the $a_-$ optical mode. 
Due to the different optical susceptibility of the forward and backward subsystems for any non-zero optomechanical coupling, the system exhibits broken time-reversal symmetry for transmission measurements.
For reflections to take place in this system, i.e., Port 1 $\rightarrow$ Port 1, or Port 2 $\rightarrow$ Port 2, light must interact in series with both the forward and backward subsystems while passing through the Rayleigh backscattering channel (see illustration in Supplement Fig.~S1). Thus, the reflection coefficients measured at each port are necessarily identical.

We can analytically obtain the waveguide transmission coefficients ($T_{21}$ in the forward direction, $T_{12}$ in the backward direction) and reflection coefficients at each port ($R_{11} = R_{22} = R$) using the Heisenberg-Langevin equations for motion for this system in the rotating wave approximation (Supplement \S S1).
In any side coupled resonator-waveguide system, the total optical loss rate $\kappa$ is defined by both the extrinsic losses (from waveguide loading) and intrinsic losses through the expression $\kappa = \kappa_i + \kappa_{\text{ex}}$. 
The condition for `critical coupling' -- defined as the point where on-resonance transmission reaches zero -- can be derived as $\kappa = 2 \kappa_{\text{ex}}$. Typically, this situation occurs when the intrinsic coupling rate $\kappa_{\text{ex}}$ and the intrinsic loss rate $\kappa_i$ are matched.
However in the case where both Rayleigh backscattering $V$ and the optomechanical coupling $G$ 
are acting on the modes (Fig.~\ref{fig:1}d), the optical loss rates for the $a_\pm$ modes are no longer identical and the critical coupling conditions must also change. 
In the simplest case in which all fields are on-resonance, the total effective loss rates for the $a_\pm$ modes can be evaluated as (details in Supplement \S S1.2) :
\begin{subequations}
	\begin{align}
	 \kappa_{\text{eff}}^+ &= \kappa \left( 1+ \mathcal{C} \right) +   \frac{4V^2}{\kappa} \label{eq:k_eff_cw}\\
	 \kappa_{\text{eff}}^- &= \kappa + \frac{4V^2}{\kappa (1+ \mathcal{C})} ~. \label{eq:k_eff_ccw}
	\end{align}
	\label{eq:k_eff}%   the extra % after the label fixes the indentation problem for the next para
\end{subequations}
Here we have introduced $\mathcal{C} = 4G^2/\kappa \Gamma$ as the optomechanical cooperativity.
These expressions show that even if $\mathcal{C} = 0$ the Rayleigh backscattering introduces additional intrinsic optical loss of $4 V^2/\kappa$ to each mode, a loss channel formed through the counterpropagating mode.
Moreover, we can see that as the optomechanical coupling rate is increased in the cw direction, light on-resonance in the \textit{ccw} mode experiences a reduction in optical loss ($\kappa_{\text{eff}}^-$ reduces) as an indirect effect.
In the limit of large optomechanical coupling $\mathcal{C}\rightarrow\infty$ the optical loss in the ccw mode approaches $\kappa$, i.e., the effective intrinsic loss $\kappa_{\text{eff}}^- - \kappa_{\text{ex}}$ approaches the purely intrinsic loss rate $\kappa_i$.
In this case the reflection should also approach zero since no backscattering occurs.

%%%%%%%%%%%%%%%%%%%%%%%%%%
%% Experimental methods
%%%%%%%%%%%%%%%%%%%%%%%%%%

\vspace{12pt}

\begin{figure}[ph]
	\begin{adjustwidth}{-1in}{-1in}
		\vspace{-2cm} 
		\makebox[\textwidth][c]{\includegraphics[width=1.15\textwidth]{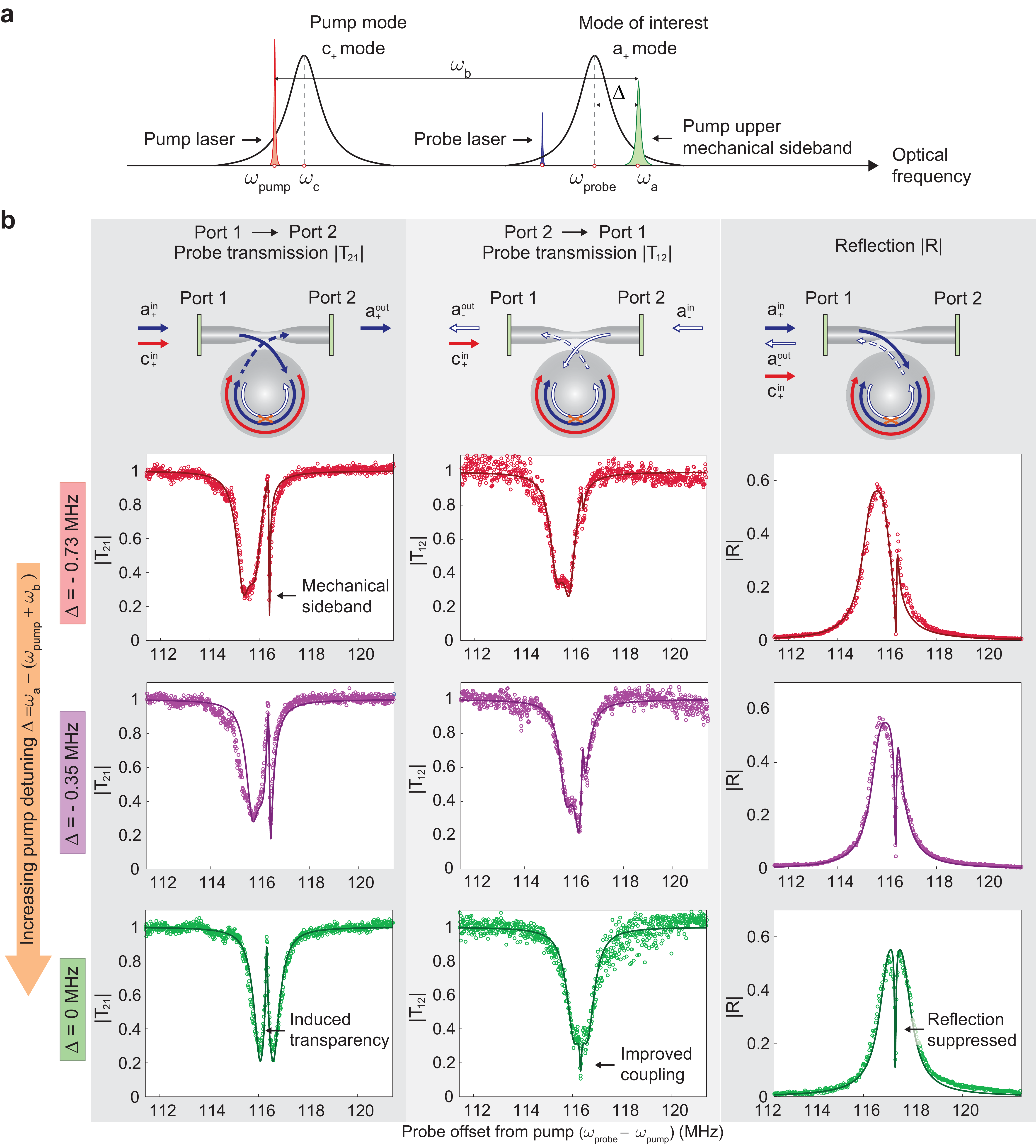}}
		\centering
		\caption{
			\textbf{Demonstration of dynamic optomechanical suppression of Rayleigh backscattering.} 
			\textbf{(a)} General configuration of optical pump, probe, and mechanical sidebands with respect to the $c_+$ and $a_+$ optical modes in the cw direction -- used throughout this work.
			\textbf{(b)}
			This experiment uses a 116 MHz mechanical mode in a 90 $\mu$m radius silica WGR. The Rayleigh scattering induced doublet is readily observed in probing of the $a_\pm$ optical modes for an off-resonance pump.
			As the pump is brought on resonance, the Brillouin scattering induced transparency is generated for the cw mode only (its on-resonance susceptibility is reduced), which breaks time-reversal symmetry within the bandwidth of the $a_\pm$ modes. 
			Observations confirm two key predictions of the model -- reduced optical reflection and improved coupling of the $a_-$ mode to the waveguide -- both confirming the suppression of Rayleigh backscattering within the WGR.
			Solid lines are simultaneous fits to the theoretical model.
			The reflection is very large due to resonant enhancement.
		}
		\label{fig:3}
	\end{adjustwidth}
\end{figure}

In order to test the predictions of the theoretical model, we perform a series of experiments using silica microsphere WGRs.
In our first experiment with a $90 \,\mu$m radius resonator, the BSIT interaction is mediated by a mechanical whispering gallery mode of frequency $\omega_b = 116.3$~MHz, having azimuthal order M = 17 around the resonator equator (corresponding to phonon wavenumber $0.030\,\upmu m^{-1}$) and mechanical damping rate $\Gamma = (14.4\pm0.2)$ kHz. 
Experimental measurements of the optical doublet (normal mode splitting) resulting from Rayleigh backscattering in this device, without any optomechanical influence, were previously shown in Fig.~\ref{fig:1}b.
The pump (for controlling $G$) and probe (to observe $a_\pm$) are produced using a 1550 nm tunable external cavity diode laser and are evanescently coupled to the optical modes using a tapered fiber waveguide as illustrated in Fig.~\ref{fig:1}a. Further details on the experimental setup and calibration of the transmission and reflection coefficients are provided in the Supplement \S S2.
In Fig.~\ref{fig:3} we present the measured transmission and reflection coefficients vs relative detuning $\Delta = \omega_a - (\omega_\text{pump} + \omega_b)$ between the $a_+$ optical mode and the pump laser anti-Stokes mechanical sideband. 
The BSIT transparency window in the forward subsystem can be clearly observed as the relative detuning $\Delta$ approaches zero.
Simultaneous measurements of the backward subsystem -- a direction in which no pumping is performed -- show that the $a_-$ mode moves closer to critical coupling within the bandwidth of the non-reciprocal effect.
Measurements of the reflection coefficient $|R|$ independently confirm the suppression of Rayleigh backscattering within the WGR.
The detailed model presented in the Supplement allows extraction of the intrinsic optical loss rate $\kappa_i = (0.35 \pm 0.03)$~MHz, the extrinsic optical loss rate $\kappa_{\text{ex}}=(0.54\pm0.01)$~MHz (the optical mode is over-coupled), the Rayleigh backscattering rate $V=(0.34\pm0.01)$~MHz, for this experiment by means of simultaneous fitting of all the measured traces. At resonance ($\Delta = 0$~MHz) we estimate $G = (0.17\pm0.01)$~MHz with an intracavity (pump) occupation number of $n_{c+} \simeq 1\times 10^{10}$ and single photon optomechanical coupling rate $g_o=(1.6\pm0.09)$~Hz. 
All uncertainties in this manuscript correspond to 95 \% confidence intervals of the fitted value.

\vspace{12pt}

\begin{figure}[tp!]
	\begin{adjustwidth}{-1in}{-1in}
	\makebox[\textwidth][c]{\includegraphics[width=1.4\textwidth]{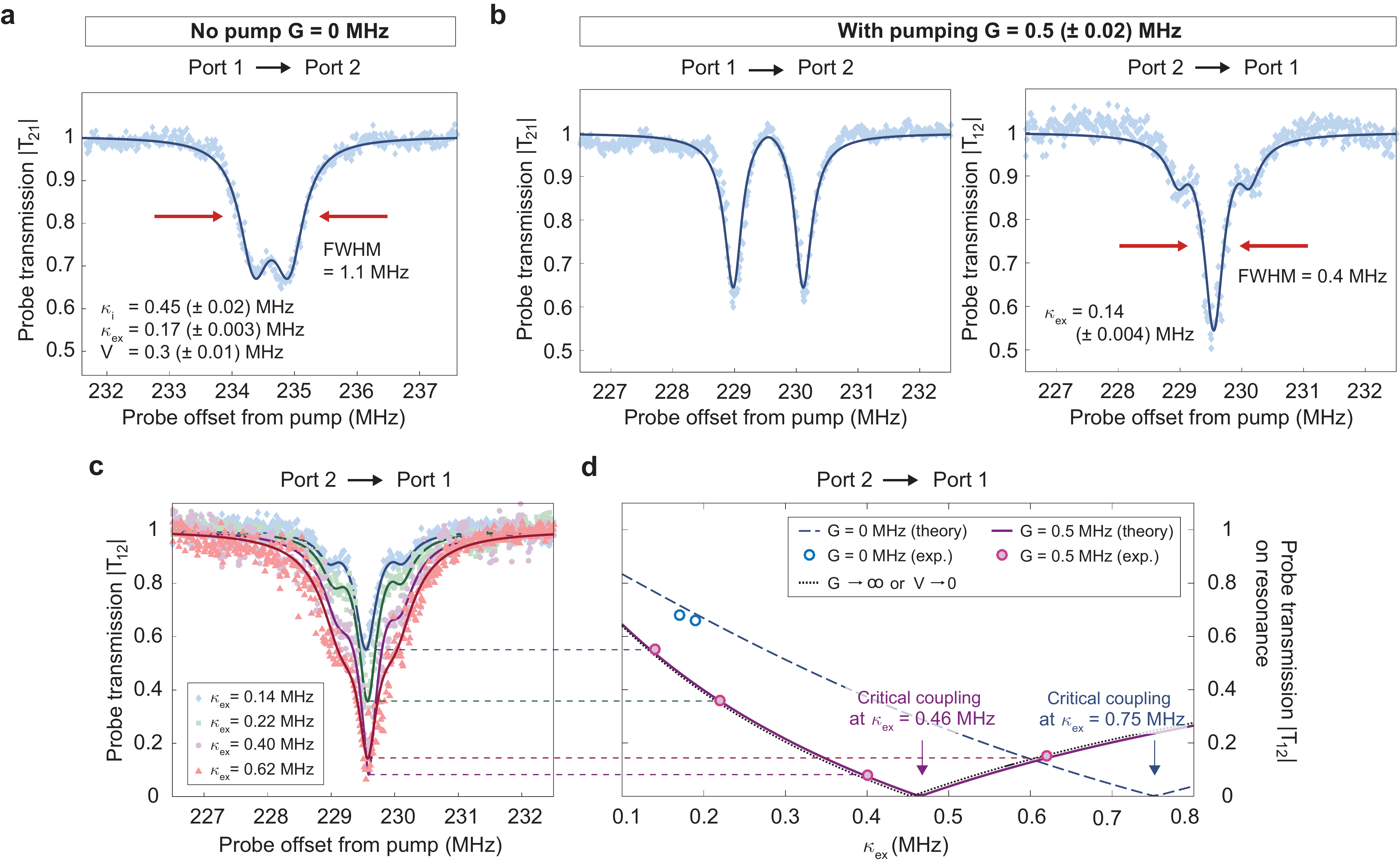}}
	\centering
	\caption{
		\textbf{Near-complete suppression of Rayleigh backscattering.} 
		\textbf{(a)} This experiment was performed with a 229.5 MHz mechanical mode of a 101 $\mu$m radius silica WGR. By initially detuning the pump laser we are able to observe the Rayleigh backscattering induced optical doublet without optomechanical pumping. Fitting to the theoretical model (solid line) indicates intrinsic loss $\kappa_i$, extrinsic loss $\kappa_{\text{ex}}$, and backscattering rate $V$.
		\textbf{(b)} We now tune the pump to obtain strong cw optomechanical coupling (with $G=0.5$ MHz), resulting in prominent change of susceptibility for the cw ($a_+$) mode only. The time-reversed ($a_-$) mode, which we did not modify, simultaneously exhibits much narrower linewidth, and the scattering induced doublet is eliminated.
		\textbf{(c)} By adjusting the extrinsic coupling to the waveguide $\kappa_{\text{ex}}$, we are able to explore the point of critical coupling for the ccw mode $a_-$.
		\textbf{(d)} The measured on-resonance transmission from experimental measurements in (c) are well matched to the theoretical predictions from the model (Supplement Eqn.~S8). With zero optomechanical coupling the critical coupling point indicates intrinsic optical loss rate of $0.75$~MHz. 
		However, with optomechanical coupling of $G = 0.5$ MHz we estimate a ccw effective intrinsic loss rate of $0.46$ MHz, a very close match to the purely intrinsic loss rate of $0.45$ MHz.
		} 
	\label{fig:4}
	\end{adjustwidth}
\end{figure}

To explore near-complete suppression of Rayleigh backscattering, we performed a second experiment on a 101 $\mu$m radius resonator (Fig.~\ref{fig:4}), mediated by a $229.5$ MHz mechanical mode having azimuthal order M = 39 (corresponding to phonon wavenumber $0.062 \,\upmu m^{-1}$) and damping rate $\Gamma = (39.1\pm1.3)$ kHz. 
Here, we used higher pump power to bring the system into the normal-mode coupling regime~\cite{Kim:isolator, Groblacher:2009}, where the optomechanical coupling exceeds the total optical loss rate ($G \geq \kappa/2$).
We first detune the pump sufficiently so that the optomechanical coupling is negligible (Fig.~\ref{fig:4}a). 
The $a_\pm$ modes are seen to hybridize and exhibit the doublet characteristic as expected from Rayleigh backscattering within the resonator. 
Fitting to the theoretical model allows us to discern the backscattering strength at $V=(0.3 \pm 0.01)$~MHz, intrinsic loss $\kappa_i = (0.45 \pm 0.02)$~MHz, and extrinsic coupling for the experiment at $\kappa_{\text{ex}}=(0.17 \pm 0.003)$~MHz, all of which contribute to the measured lineshape. 
We now bring the cw pump laser on resonance ($\Delta = 0$~MHz) such that $272~\mu$W optical power is absorbed into the resonator, leading to an estimated optomechanical coupling rate $G=(0.5 \pm 0.02)$~MHz. 
The resulting optomechanically induced normal mode splitting in the $a_+$ mode can be observed through forward transmission measurement (Fig.~\ref{fig:4}b-left). 
Since the TRS broken bandwidth now encompasses nearly the whole of the counter-propagating $a_-$ mode, the backscattering is almost completely eliminated. 
Simultaneous measurement of backward transmission reveals that the Rayleigh backscattering induced doublet disappears in the $a_-$ mode, and there is a significant improvement in its linewidth. Further, the on-resonance transmission dips lower indicating that coupling with the waveguide is closer to critical (Fig.~\ref{fig:4}b-right).

As discussed above (Eqn.~\ref{eq:k_eff}b), in the limit of large optomechanical coupling the Rayleigh backscattering can be mitigated completely and the effective intrinsic loss of the ccw $a_-$ mode must converge to its purely intrinsic loss rate $\kappa_i = 0.45$~MHz.
In such a case, we would expect to see critical coupling in the waveguide-resonator system when extrinsic coupling $\kappa_{\text{ex}} = 0.45$ MHz. 
In fact, measurement of this critical coupling point is the only directly accessible measurement of the on-resonance intrinsic optical loss since the optical modeshape is non-Lorentzian.
We test this complete suppression scenario by subsequently increasing the extrinsic waveguide-resonator coupling from 0.14 MHz to 0.62 MHz.
Experimental measurements of the ccw mode in Fig.~\ref{fig:4}c and \ref{fig:4}d show the evolution of the ccw resonance, as it proceeds from undercoupling to overcoupling while passing through the critical coupling point. 
Data for zero optomechanical coupling are well matched to the theoretical predictions, showing that critical coupling occurs at $\kappa_{\text{ex}} = 0.75$~MHz due to the increased intrinsic loss ($+4 V^2 / \kappa$) from the Rayleigh backscattering channel.
On the other hand, $G=0.5$ MHz data and theoretical prediction show that critical coupling must occur at $\kappa_{\text{ex}} \approx 0.46$ MHz. This is extremely close to the estimated intrinsic loss rate of $\kappa_i = 0.45$ MHz indicating that the TRS broken system achieves nearly complete suppression of the undesirable Rayleigh backscattering.
As predicted by the theoretical model (Supplement \S S1.4) an even higher optomechanical coupling rate could push the `optomechanical wings' further out, permitting recovery of the Lorentzian lineshape of the ccw optical mode.

\vspace{12pt}

The approach that we demonstrate here for dynamic suppression of Rayleigh backscattering employs a local time-reversal symmetry breaking interaction, i.e., Brillouin scattering, that is available in all phases of matter. 
Due to the ubiquity of this effect, our system can in principle be unwrapped for linear waveguide systems as well, where immunity against unforeseen defects is highly desirable. For example, an optical waveguide could be made robust against backscattering from a damaged segment by suppressing the time-reversed photonic density of states through the optomechanical influence.
More broadly, this principle of suppressing backscattering via TRS breaking may also be readily achieved through other local nonreciprocal techniques, encompassing nonlinear optics 
	\cite{Kang:2011, Shen:2016}, 
chirally pumped atoms \cite{Sayrin:2015}, parity-time symmetry breaking \cite{Rueter:2010, Peng:2014}, and spatiotemporal modulation 
	\cite{Lira:2012}.

%\end{linenumbers}

\newpage
\bibliographystyle{naturemag}
\bibliography{bib_rayleigh}

\end{document}

\title{Supplementary information:\\
	 Dynamic suppression of Rayleigh light scattering \\ in dielectric resonators}

\author{Seunghwi Kim$^1$, Jacob M. Taylor$^{2,3}$, and Gaurav Bahl$^{1\ast}$\\
	\\
	\footnotesize{$^1$ Mechanical Science and Engineering, University of Illinois at Urbana-Champaign,}\\
	\footnotesize{Urbana, Illinois 61801, USA}\\
	\footnotesize{$^2$ Joint Quantum Institute, University of Maryland, }\\
	\footnotesize{College Park, Maryland 20742, USA}\\
	\footnotesize{$^3$ Joint Center for Quantum Information and Computer Science,} \\
	\footnotesize{National Institute of Standards and Technology, Gaithersburg, Maryland 20899, USA}\\
	\footnotesize{$^\ast$ To whom correspondence should be addressed; bahl@illinois.edu} \\
}

	\maketitle
%	\date{}

	\tableofcontents
	
	\newpage

		\newpage

%%%%%%%%%%%%%%%%%%%%%%%%%%%%%%%%
\section{System model including Rayleigh scattering and optomechanical coupling} 
%%%%%%%%%%%%%%%%%%%%%%%%%%%%%%%%

Our whispering gallery resonator (WGR) system supports two frequency-adjacent optical modes, the mode of interest $a_\pm$ and Stokes shifted pump mode $c_\pm$, and a mechanical mode $b_\pm$. All three modes are of whispering gallery mode (WGM) type and exist as degenerate pairs in the cw (+) and ccw (-) direction. As described in the main text (Fig.~1), the cw optical modes $a_+$ and $c_+$ couple through the cw mechanical mode $b_+$ via optomechanical interaction:
\begin{align*}
	H_{\text{int}}^{\text{OM}} = \hbar(g_o c_+ a_+^{\dagger} b_+ + g_o^* c_+^{\dagger} a_+ b^{\dagger}_+)
\end{align*} 
We define the single photon Brillouin optomechanical coupling rate $g_o \propto  \delta(\Delta k)\int \phi_1 \phi_2 \psi \, d^2r$, where $\phi_1$, $\phi_2$ and $\psi$ are the transverse mode shapes of the optical and mechanical modes, respectively. The delta function $\delta(\Delta k)$ represents the momentum selection condition for Brillouin scattering, i.e. the momentum difference between the optical modes $a_+$ and $c_+$ must match the momentum of the mechanical mode $b_+$. 
We also consider the interaction of the cw optical modes ($a_+$, $c_+$) with their time-reversed counterparts ($a_-$, $c_-$) via elastic Rayleigh backscattering. This coupling can be induced by surface or internal inhomogeneities~\cite{Weiss:95, Gorodetsky:00, Kippenberg:2002, Matsko:2017} in WGRs.
Under the dipole approximation, we can write the interaction Hamiltonian due to Rayleigh scattering for the optical mode pairs as follows \cite{Mazzei2007}:
\begin{align*}
	H_{\text{int}}^{\text{R}} = \hbar V_o (a_+^{\dagger} a_-  + a_-^{\dagger} a_+) +\hbar V_1 (c_+^{\dagger} c_- + c_-^{\dagger} c_+)
\end{align*}
Here, we have defined $V_o$ and $V_1$ as the backscattering rates for the $a_\pm$ and $c_\pm$ modes, respectively. These coupling rates are given by $2V_{i=0,1} = -\alpha f_i^2(r) \omega_i / \mathcal{V}_{\text{m}}^i $ where $\alpha$ is the polarizability of the scatterer, $f_i(r)$ accounts for the overlap of the optical field with the scatterer dipole, $\omega_i$ is the resonant frequency of the optical mode and $\mathcal{V}_{\text{m}}^i$ is its modal volume~\cite{Mazzei2007}. The normal-mode splitting induced by Rayleigh backscattering is easily experimentally observable if $V_i > \kappa/2$.

Considering the two interaction Hamiltonians, we can now represent the linearized Heisenberg-Langevin equations of our system. Under the non-depleted pump approximation we are able to omit the equations for $c_\pm$, which leads to the equations of motion:
\begin{subequations}
\begin{align}
	\frac{d a_+}{dt}  &= -\left( \frac{\kappa}{2} + i\Delta_a \right) a_+ - iGb_+ - iV_o a_- + \sqrt{\kappa_{\text{ex}}}a_+^{\text{in}}(t) + \sqrt{\kappa_i}a_\text{vac}(t),\\
	\frac{d b_+}{dt} &= -\left( \frac{\Gamma}{2} + i\Delta_b \right) b_+ -iG^*a_+ +\sqrt{\Gamma} b_{\text{th}}(t),\\
	\frac{d a_-}{dt} &= -\left( \frac{\kappa}{2} + i\Delta_a \right) a_- -  \eta Gb_- - iV_o a_+ + \sqrt{\kappa_{\text{ex}}}a_-^{\text{in}}(t) + \sqrt{\kappa_i}a_\text{vac}(t),\\
	\frac{d b_-}{dt} &= -\left( \frac{\Gamma}{2} + i\Delta_b \right) b_- +\eta G^*a_- +\sqrt{\Gamma} b_{\text{th}}(t).				
\end{align}
\label{eq:eom}
\end{subequations}

\noindent
Here $a_{\pm}^{\text{in}}$ are the normalized probe laser amplitudes within the waveguide, in forward (+) and backward (-) directions. They are defined as  $|a_{\pm}^{\text{in}}|^2 = P_{\text{probe}\pm}^{\text{in}}/ \hbar \omega_{\text{probe}}$, where $P_{\text{probe}\pm}^{\text{in}}$ is the corresponding input probe laser power into the waveguide and $\omega_{\text{probe}}$ is the probe laser frequency. $\sqrt{\kappa_{\text{ex}}}$ appears due to the external coupling to a side-coupled waveguide. $a_\text{vac}(t)$ and $b_{\text{th}}(t)$ are the vacuum and thermal noise in the optical and mechanical modes, respectively. $\kappa$ and $\Gamma$ are the total loss rates of the $a_{\pm}$ and $b_{\pm}$ modes respectively. The detuning terms are defined as $\Delta_a = \omega_a - \omega_{\text{probe}}$ and $\Delta_b = \omega_b - (\omega_{\text{probe}} - \omega_{\text{pump}})$, where $\omega_a$ and $\omega_b$ are resonant frequencies of $a_\pm$ and $b_\pm$ modes respectively, and $\omega_{\text{pump}}$ is the pump laser frequency. 
$G \triangleq g_o \sqrt{n_p}$ is the pump-enhanced optomechanical coupling due to the $c_+$ pump. Since we also wish to take into account Rayleigh scattering for the $c_\pm$ modes, there may be some backscattered pump power from the cw pump mode $c_+$ into the ccw pump mode $c_-$, which creates non-zero optomechanical interaction in the ccw direction. In the above equations we have incorporated this ccw optomechanical interaction by introducing $\eta = 2V_1/ \kappa_c$, where $\kappa_c$ is the optical loss rate of the $c_\pm$ modes.  
After this accounting we are no longer interested in the pump equations of motion, so we can dispense with the $V_{0,1}$ distinctions and instead replace a single backscattering rate $V = V_0$ between the $a_\pm$ modes.

\vspace{12pt}

\begin{figure}[hp]
	\begin{adjustwidth}{-1in}{-1in}	
		\makebox[\textwidth][c]{\includegraphics[width=0.94\textwidth]{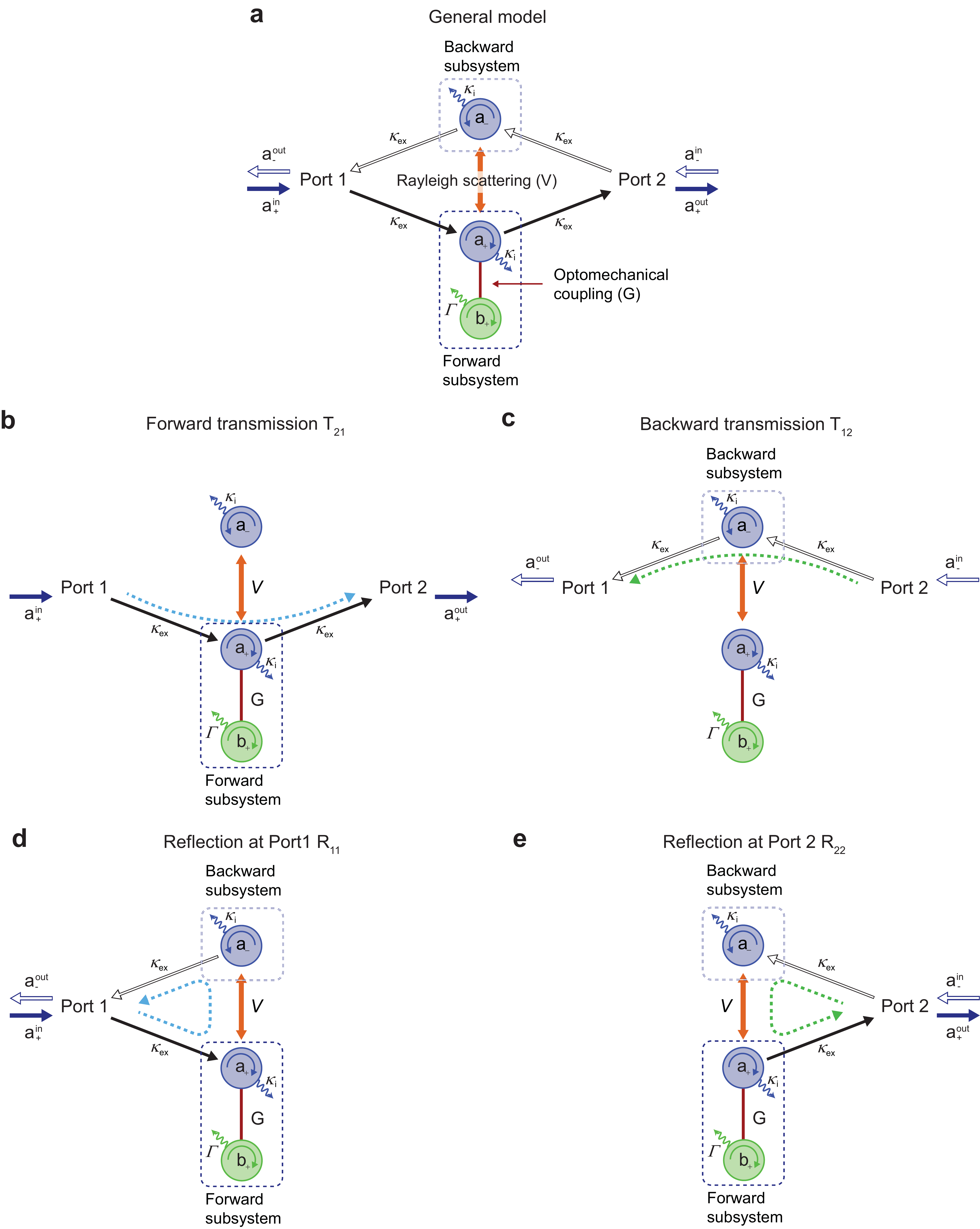}}
		\centering
		\caption{
			\textbf{Toy model for transmission and reflection coefficients.} 
			\textbf{(a)} Our system can be described through a two-port system picture, as described in the main manuscript, where each port indicates the left or right ends of the waveguide. Using this, we can describe \textbf{(b)} the forward transmission coefficient ($T_{21}$), \textbf{(c)} the backward transmission coefficient ($T_{12}$), \textbf{(d)} the reflection coefficient at Port 1 ($R_{11}$), and \textbf{(e)} the reflection coefficient at Port 2 ($R_{22}$). 
			The two reflection coefficients $R_{11}$ and $R_{22}$ must be always identical since the interaction takes place through both forward and backward subsystems.
		}
		\label{fig:model_detail}
	\end{adjustwidth}
\end{figure}

Fig.~\ref{fig:model_detail} presents the toy model of our system, in which the forward and backward subsystems of the resonator are deliberately identified separately. As explained in the main text, the transmission coefficients must be different (Fig.~\ref{fig:model_detail}b and \ref{fig:model_detail}c). On the other hand, the two reflection coefficients must be identical since the light experiences both forward and backward optical susceptibilities in series (Fig.~\ref{fig:model_detail}d and \ref{fig:model_detail}e), i.e. the reflection system is identical in either direction ($R=R_{11} = R_{22}$).

\FloatBarrier

%\vspace{24pt}

%%%%%%%%%%%%%%%%%%%%%%%%%%%%%%%%
\subsection{Waveguide transmission and reflection coefficients} 
%%%%%%%%%%%%%%%%%%%%%%%%%%%%%%%%

To experimentally investigate the optomechanical modification of Rayleigh backscattering within the resonator, we can perform measurements of the transmission and reflection coefficients through the side-coupled waveguide. Since we are interested in the stationary solutions of Eqns.~\ref{eq:eom}, we can neglect the vacuum and thermal noise in our calculation. The steady state intracavity field solutions $\bar{a}_+$ and $\bar{a}_-$ excited by both forward and backward probe fields are obtained as follows : 
\begin{subequations}
	\begin{align}
	\bar{a}_+ &= \ddfrac{\sqrt{k_{\text{ex}}} a_+^{\text{in}} -\ddfrac{iV ( \Gamma /2 + i \Delta_b) \sqrt{\kappa_{\text{ex}}} a_-^{\text{in}}}{(\kappa / 2 + i \Delta_a)( \Gamma /2 + i \Delta_b) + \eta^2 G^2}}{ \ddfrac{\kappa}{2} + i \Delta_a + \ddfrac{G^2}{\Gamma/2 + i \Delta_b} +\ddfrac{V^2(\Gamma / 2 + i \Delta_b)}{(\kappa / 2 + i \Delta_a)(\Gamma / 2 + i \Delta_b) +\eta^2 G^2 }}, \label{eq:a_cw}\\
	\bar{a}_- &= \ddfrac{\sqrt{k_{\text{ex}}} a_-^{\text{in}} - \ddfrac{ i V  ( \Gamma /2 + i \Delta_b) \sqrt{\kappa_{\text{ex}}} a_+^{\text{in}} }{(\kappa / 2 + i \Delta_a)  ( \Gamma /2 + i \Delta_b) + G^2}}{ \ddfrac{\kappa}{2} + i \Delta_a + \frac{\eta^2 G^2}{\Gamma / 2 + i \Delta_b} + \ddfrac{V^2 ( \Gamma /2 + i \Delta_b)}{(\kappa / 2 + i \Delta_a) ( \Gamma /2 + i \Delta_b) + G^2} }. \label{eq:a_ccw}
	\end{align}
	\label{eq:steady_sol}
\end{subequations}

\begin{figure}[t]
	\begin{adjustwidth}{-1in}{-1in}	
	\makebox[\textwidth][c]{\includegraphics[width=0.8\textwidth]{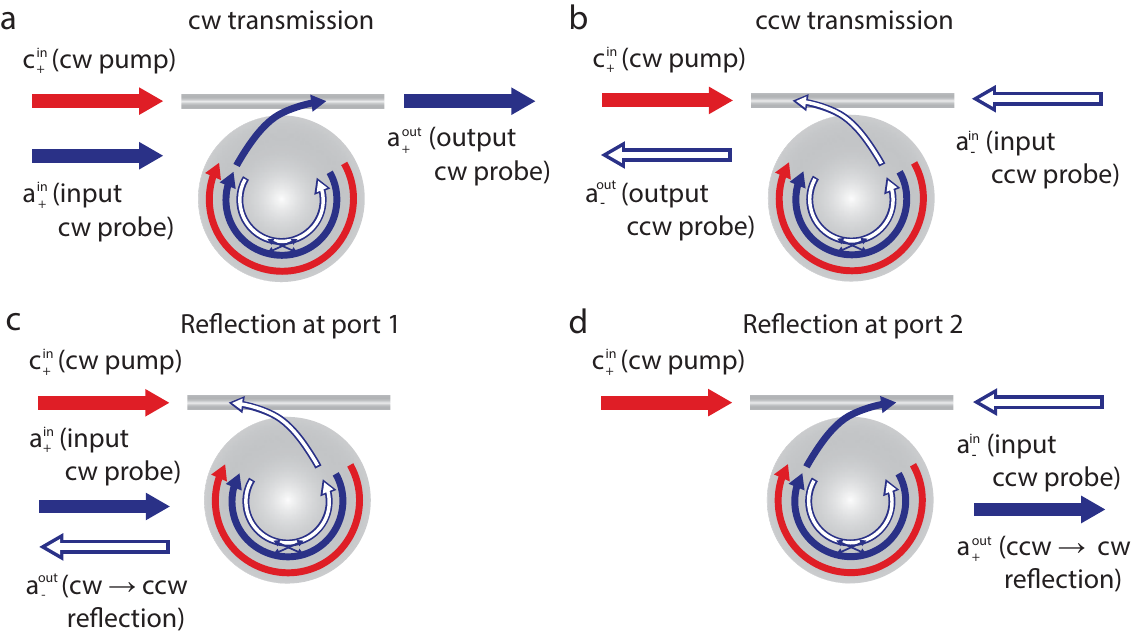}}
	\centering
	\caption{\textbf{Variable descriptions for transmission and reflection measurement.} For cw optical pumping into the $c_+$ mode, we can define \textbf{(a)} forward transmission, \textbf{(b)} backward transmission, \textbf{(c)} reflection of the input cw probe $a_+^{\text{in}}$ within the probe optical modes $a_\pm$, and \textbf{(d)} reflection of the input ccw probe $a_-^{\text{in}}$ within the probe optical modes $a_\pm$. As explained in Fig.~\ref{fig:model_detail}d and \ref{fig:model_detail}e the reflections identified in (c) and (d) must be identical.}
	\label{fig:3_cases}
	\end{adjustwidth}
\end{figure}

The above expressions show that the cavity modes can be populated by both forward and backward optical probes, due to the Rayleigh backscattering. Using the resonator input-output formalism, we now can obtain expressions for the output fields in the waveguide (Fig.~\ref{fig:3_cases}).
\begin{compactenum}
	\item for cw transmission $\left. a_+^{\text{out}} \right\rvert_{a_-^{\text{in}} = 0} = \left. a_+^{\text{in}} - \sqrt{\kappa_{\text{ex}}}\,\bar{a}_+ \right\rvert_{a_-^{\text{in}} = 0}$
	\item for ccw transmission $\left. a_-^{\text{out}} \right\rvert_{a_+^{\text{in}} = 0} =\left. a_-^{\text{in}} - \sqrt{\kappa_{\text{ex}}}\,\bar{a}_-\right\rvert_{a_+^{\text{in}} = 0}$
	\item for cw $\rightarrow$ ccw reflection $\left. a_-^{\text{out}} \right\rvert_{a_-^{\text{in}} = 0} =  -\left.\sqrt{\kappa_{\text{ex}}}\,\bar{a}_- \right\rvert_{a_-^{\text{in}} = 0}$
	\item for ccw $\rightarrow$ cw reflection $\left. a_+^{\text{out}} \right\rvert_{a_+^{\text{in}} = 0} = -\left.\sqrt{\kappa_{\text{ex}}}\,\bar{a}_+ \right\rvert_{a_+^{\text{in}} = 0}$
\end{compactenum}

\vspace{12pt}

The above expressions allow us to derive the waveguide transmission and reflection coefficients as follows :
\begin{subequations}
\begin{align}
	T_{21} &= \left.\ddfrac{a_+^{\text{out}}}{a_+^{\text{in}}} \right\rvert_{a_-^{\text{in}} = 0}  = 1 - \ddfrac{\kappa_{\text{ex}}}{ \ddfrac{\kappa}{2} + i \Delta_a + \ddfrac{G^2}{\Gamma/2 + i \Delta_b} +\ddfrac{V^2(\Gamma / 2 + i \Delta_b)}{(\kappa / 2 + i \Delta_a)(\Gamma / 2 + i \Delta_b) +\eta^2 G^2 }},\\
	T_{12} &= \left.\ddfrac{a_-^{\text{out}}}{a_-^{\text{in}}} \right\rvert_{a_+^{\text{in}} = 0} = 1 - \ddfrac{\kappa_{\text{ex}}}{ \ddfrac{\kappa}{2} + i \Delta_a + \ddfrac{V^2}{\kappa / 2 + i \Delta_a + G^2/ ( \Gamma /2 + i \Delta_b)} + \frac{\eta^2 G^2}{\Gamma / 2 + i \Delta_b}},\label{eq:t_ccw}\\
	R & = \left.\ddfrac{a_-^{\text{out}}}{a_+^{\text{in}}}\right\rvert_{a_-^{\text{in}} = 0} = \left.\ddfrac{a_+^{\text{out}}}{a_-^{\text{in}}}\right\rvert_{a_+^{\text{in}} = 0} = \ddfrac{ \ddfrac{ i V \kappa_{\text{ex}} }{\kappa / 2 + i \Delta_a + G^2/ ( \Gamma /2 + i \Delta_b)}}{ \ddfrac{\kappa}{2} + i \Delta_a + \ddfrac{V^2}{\kappa / 2 + i \Delta_a + G^2/ ( \Gamma /2 + i \Delta_b)} + \frac{\eta^2 G^2}{\Gamma / 2 + i \Delta_b}}. 
\end{align}
\label{eq:T_R}
\end{subequations}

%%%%%% MOVED IN FROM MAIN TEXT %%%%%%%%%%%
We can also consider the case where there is no backscattering of the pump, i.e. $\eta = 0$, which results in the simplified equations :
\begin{subequations}
	\begin{align}
	T_{21} &= \left.\ddfrac{a_+^{\text{out}}}{a_+^{\text{in}}} \right\rvert_{a_-^{\text{in}} = 0} = 1 - \ddfrac{\kappa_{\text{ex}}}{ \ddfrac{\kappa}{2} + i \Delta_a + \ddfrac{G^2}{\Gamma/2 + i \Delta_b} +\frac{V^2}{\kappa / 2 + i \Delta_a}} \label{eq:t_cw}\\
	T_{12} &= \left.\ddfrac{a_-^{\text{out}}}{a_-^{\text{in}}} \right\rvert_{a_+^{\text{in}} = 0} = 1 - \ddfrac{\kappa_{\text{ex}}}{ \ddfrac{\kappa}{2} + i \Delta_a + \ddfrac{V^2}{\kappa / 2 + i \Delta_a + G^2/ ( \Gamma /2 + i \Delta_b)}}\label{eq:t_ccw_add}\\
	R &  = \left.\ddfrac{a_{-}^{\text{out}}}{a_+^{\text{in}}}\right\rvert_{a_-^{\text{in}} = 0} = \left.\ddfrac{a_{+}^{\text{out}}}{a_-^{\text{in}}}\right\rvert_{a_+^{\text{in}} = 0} = \ddfrac{ \ddfrac{ i V \kappa_{\text{ex}} }{\kappa / 2 + i \Delta_a + G^2/ ( \Gamma /2 + i \Delta_b)}}{ \ddfrac{\kappa}{2} + i \Delta_a + \frac{V^2}{\kappa / 2 + i \Delta_a + G^2/ ( \Gamma /2 + i \Delta_b)}} \label{eq:ref}
	\end{align}\label{eq:coeff}
\end{subequations}
%%%%%% MOVED IN FROM MAIN TEXT %%%%%%%%%%%

Finally, we can also produce a scattering matrix formalism for the the optical probe transmission and reflection coefficients, that incorporates simultaneous inputs from both directions in the waveguide: 
\begin{align}
	\begin{pmatrix}
		a_-^{\text{out}} \\
		a_+^{\text{out}}
	\end{pmatrix}
	=
	\begin{pmatrix}
		R_{11} & T_{12}  \\
		T_{21} & R_{22}
	\end{pmatrix}
	\begin{pmatrix}
		a_+^{\text{in}} \\
		a_-^{\text{in}}
	\end{pmatrix} \label{eq:s_matrix}
\end{align}
Here we see that the scattering matrix is generally non-symmetric, i.e. $T_{12} \neq T_{21}$, when the optomechanical coupling is non-zero.
This non-reciprocity of transmission coefficients induced through Brillouin scattering has been already reported~\cite{Kim:2015, Dong:2015}.

\vspace{12pt}

%%%%%%%%%%%%%%%%%%%%%%%%%%%%%%%%
\subsection{Effective optical loss rate} 
%%%%%%%%%%%%%%%%%%%%%%%%%%%%%%%%
\label{sec:losses}

Optical loss due to Rayleigh backscattering is typically included as a part of the intrinsic loss in whispering gallery resonators, since it cannot be distinguished from absorption losses at low scattering rates. 
In this work, however, we must explicitly distinguish the optical loss due to Rayleigh backscattering from other intrinsic optical losses. To quantify the optical loss, we first focus on the susceptibilities for the $a_\pm$ modes using Eqs.~\ref{eq:eom}. Solving in the Fourier domain, we obtain
\begin{subequations}
\begin{align}
	\chi_{a_+}^{-1} (\omega) &= -i(\omega -\Delta_a) + \kappa/2 \nonumber
	\\& + \frac{G^2}{-i(\omega - \Delta_b) + \Gamma/2} \nonumber 
	\\& +   \frac{V^2 (- i (\omega - \Delta_b) +\Gamma / 2 )}{(-i(\omega - \Delta_a) + \kappa/2)(- i (\omega - \Delta_b) +\Gamma / 2 ) + \eta^2 G^2},~\text{and}\\
	\chi_{a_-}^{-1} (\omega) &= -i(\omega -\Delta_a) + \kappa/2 \nonumber
	\\& +  \frac{\eta^2 G^2 }{-i (\omega- \Delta_a) +\Gamma / 2} \nonumber
	\\& + \frac{V^2 \left( -i(\omega - \Delta_b)+ \Gamma/2 \right)}{(-i(\omega - \Delta_a) + \kappa/2)  \left( -i(\omega - \Delta_b)+ \Gamma/2 \right) + G^2}. 
\end{align}
\end{subequations}
The effective optical loss, including the loss due to Rayleigh scattering and optomechanical coupling, can be extracted from real part of the optical susceptibilities. At zero detuning i.e $\Delta_a = 0$ and $\Delta_b = 0$, the total effective optical loss rates of the $a_\pm$ modes (including waveguide loading) are given by:
\begin{subequations}
	\begin{align}
	\kappa_{\text{eff}}^+ &= \kappa \left( 1+ \mathcal{C} \right) +   \frac{4V^2}{\kappa (1+ \eta^2 \mathcal{C})}, \label{eq:k_+} \\
	\kappa_{\text{eff}}^- &= \kappa(1+ \eta^2\mathcal{C}  ) + \frac{4V^2}{\kappa (1+ \mathcal{C})}  \label{eq:k_-}.
	\end{align} \label{eq:lossrates}%
\end{subequations}
where we define optomechanical cooperativity as $\mathcal{C} = 4G^2/\kappa \Gamma$. If the pump reflection ($\eta$) is small, we see that the effective loss rate of the $a_+$ mode increases with increasing $\mathcal{C}$ in Eq.~(\ref{eq:k_+}), which corresponds to the results of the optomechanically induced transparency~\cite{Weis_OMIT, Safavi-Naeini_OMIT, Kim:2015}. Meanwhile the second term in Eq.~(\ref{eq:k_-}) decreases with increasing $\mathcal{C}$. This analysis reveals that the Rayleigh backscattering contribution is effectively shut down in the limit of large $\mathcal{C}$.

\vspace{12pt}

%%%%%%%%%%%%%%%%%%%%%%%%%%%%%%%%
\subsection{Redefining the condition for critical coupling} 
%%%%%%%%%%%%%%%%%%%%%%%%%%%%%%%%
\label{sec:critical}

For conventional resonator-waveguide systems, the transmission through the waveguide in either direction is given by
\begin{align*}
	T = \frac{(\kappa - 2\kappa_{\text{ex}})/2 + i \Delta}{\kappa/2 + i \Delta}
\end{align*}
which can be derived by setting $G=$ and $V=0$ in Eqn.~\ref{eq:T_R}a.
Critical coupling, the point where on-resonance ($\Delta = 0$) transmission dips to zero in conventional resonator systems is achieved when $\kappa_{\text{ex}} = \kappa/2$. 
However this condition for achieving critical coupling must be modified in our system. 
For probing of the ccw optical mode (backward direction), we can rewrite the transmission coefficient at zero detuning ($\Delta_a=0$ and $\Delta_b=0$) as described in Eqn.~\ref{eq:t_ccw}:
\begin{align}
	T_{12} = \frac{\kappa_{\text{eff}}^- - 2\kappa_{\text{ex}}}{\kappa_{\text{eff}}^-} = \frac{\kappa(1+ \eta^2\mathcal{C}  ) + 4V^2/\kappa (1+ \mathcal{C}) - 2 \kappa_{\text{ex}}}{\kappa(1+ \eta^2\mathcal{C}  ) + 4V^2/\kappa (1+ \mathcal{C})}
\end{align}
In other words, the external coupling rate needed to reach critical coupling of the ccw $a_-$ mode with the waveguide should be modified to the following :
\begin{align}
	\kappa_{\text{ex}} = \frac{\kappa_{\text{eff}}^-}{2} =  \frac{\kappa(1+ \eta^2\mathcal{C}  )}{2} + \frac{2V^2}{\kappa (1+ \mathcal{C})}  \label{eq:c_cond}
\end{align}

\vspace{12pt} 

%%%%%%%%%%%%%%%%%%%%%%%%%%%%%%%%
\subsection{Evolution of transmission and reflection coefficients} 
%%%%%%%%%%%%%%%%%%%%%%%%%%%%%%%%

\begin{figure}[bh!]
	\begin{adjustwidth}{-1in}{-1in}
	\makebox[\textwidth][c]{\includegraphics[width=1.1\textwidth, clip=true, trim=0in 0.05in 0in 0in]{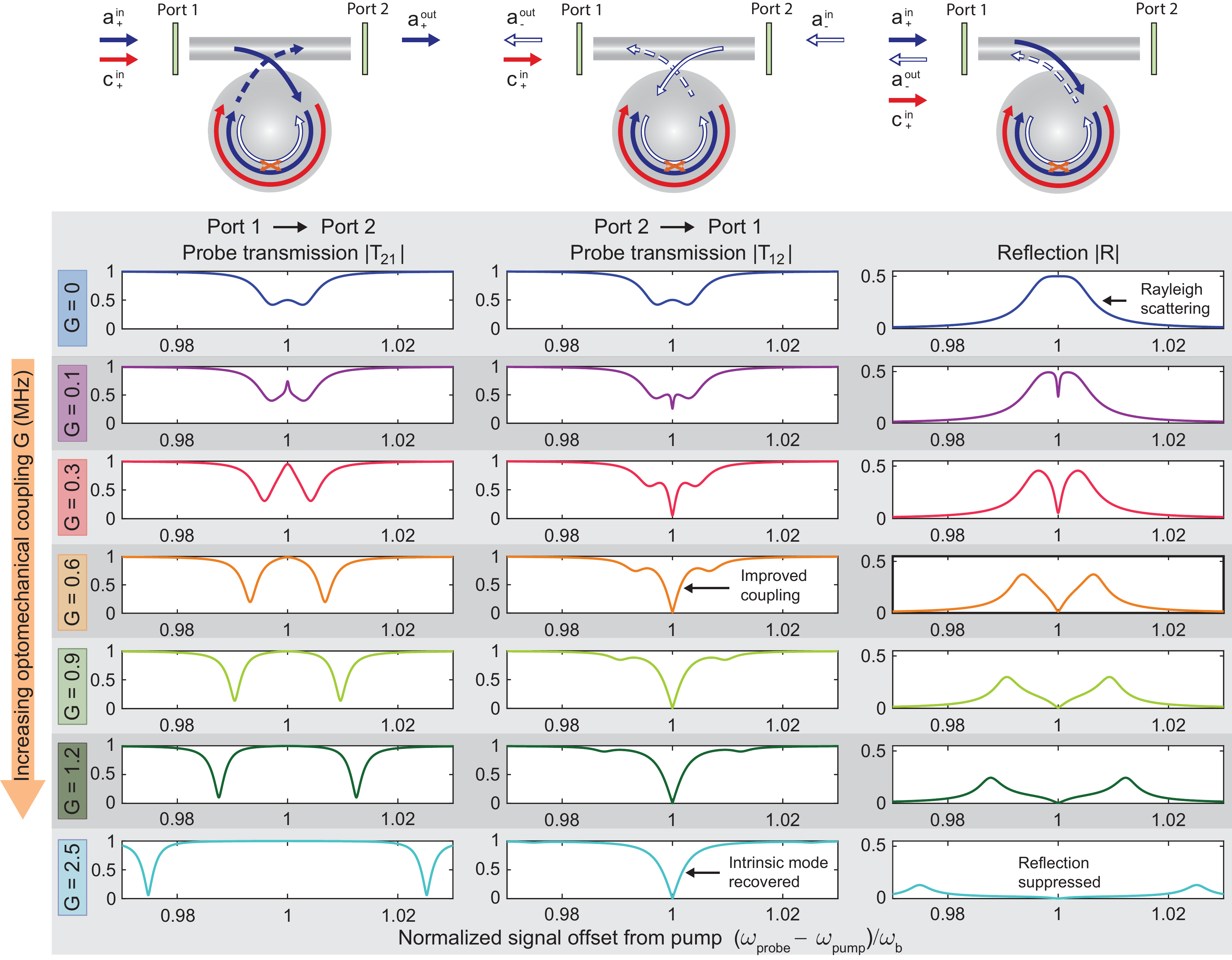}}
	\centering
	\caption{
		\textbf{Theoretical prediction of backscattering suppression.} 
		We model the waveguide transmission and reflection coefficients (Eqns.~\ref{eq:coeff}) for varying optomechanical coupling rate $G$.
		The model parameters are $\kappa = 0.7$ MHz, $\kappa_\text{ex} = 0.35$ MHz, $\Gamma = 30$ kHz, and $V = 0.35$ MHz to correspond closely with experiments below.
		Without any optomechanical coupling ($G=0$), as is typical, the $a_\pm$ modes exhibit the Rayleigh scattering induced doublet and produce identical waveguide transmission coefficients in both directions. Additionally, the resonant Rayleigh backscattering produces a large back-reflection coefficient. Since the effective optical loss for $G=0$ is $\kappa^\pm_{\text{eff}} = \kappa + 4 V^2/\kappa = 1.4$ MHz (see Eqns.~\ref{eq:lossrates}) the system is initially overcoupled at its resonance.
		However, as $G$ is increased, the time-reversal symmetry of the cw/ccw modes is broken, which can be observed through the strong distinction of transmission coefficients. The resulting suppression of Rayleigh backscattering can be seen in both the reduced reflection coefficient, as well as the improved coupling of the ccw resonator mode in the backward direction (reduced intrinsic loss). 
		}
	\label{fig:2}
	\end{adjustwidth}
\end{figure}

In Figure~\ref{fig:2} we invoke the model of Eqns.~\ref{eq:coeff} to predict the evolution of transmission and reflection coefficients as a function of optomechanical coupling rate and the optical probe detuning.
We have modeled an undercoupled situation, i.e. where the effective intrinsic loss rate of the optical modes is greater than the extrinsic loss under zero optomechanical coupling \cite{Gorodetsky:1996, Yariv:2000}, to correspond with the experiments presented in the main text.

For $G=0$ the model simply predicts the Rayleigh-scattering induced doublet of the hybridized optical modes.
When we engage the unidirectional cw pump (i.e. $G \neq 0$), the forward transmission model $T_{21}$ reveals that $a_+$ undergoes normal mode splitting caused by optomechanical coupling with the $b_+$ mechanical mode. 
%
% In other words, the BSIT reduces the on-resonance optical susceptibility for the forward subsystem as illustrated in Fig.~\ref{fig:1}d.
%
Intuitively, we anticipate that the reflection coefficient for the photons in the resonator should be reduced since the lowered photonic susceptibility of the forward subsystem `open circuits' the reflection pathway mediated by Rayleigh scattering. 
The reflection coefficient produced by the model agrees with this intuitive assertion.
More importantly, the backward transmission coefficient $T_{12}$ shows that the linewidth of the time-reversed mode $a_-$ narrows when $G$ increases, indicating that the intrinsic optical loss rate for that mode is reduced (see \S\ref{sec:losses}). A better confirmation of this reduction of intrinsic loss comes from the fact that the $a_-$ mode approaches critical coupling (zero on-resonance transmission, see \S\ref{sec:critical}) as the intrinsic loss approaches the extrinsic loss $\kappa_{ex}$.

\FloatBarrier

\vspace{24pt}

%%%%%%%%%%%%%%%%%%%%%%%%%%%%%%%%
\subsection{Condition for reaching a quantum point} 
%%%%%%%%%%%%%%%%%%%%%%%%%%%%%%%%

It is also interesting to calculate the effective temperature of the  $a_-$ optical mode due to Rayleigh backscattering. We assume $\eta =0$ for simplifying the equations of motion, and obtain the equations in the Fourier domain:
\begin{subequations}
	\begin{align}
	-i \omega \tilde{a}_+  &= - \frac{\kappa}{2} \tilde{a}_+ - iG\tilde{b}_+ - iV\tilde{a}_- + \sqrt{\kappa_i}\tilde{a}_\text{vac}(\omega)  \\
	-i \omega \tilde{b}_+ &= - \frac{\Gamma}{2}  \tilde{b}_+ -iG^*\tilde{a}_+ +  \sqrt{\Gamma} \tilde{b}_{\text{th}}(\omega)\\
	-i \omega \tilde{a}_- &= -\frac{\kappa}{2} \tilde{a}_-- iV\tilde{a}_+ + \sqrt{\kappa_i}\tilde{a}_\text{vac}(\omega)  		
	\end{align}
	\label{eq:eom_qp}
\end{subequations}
We note that optical noise for the $a_\pm$ modes is negligible due to the negligible thermal excitation of photons. In our system, the additional thermal load on the $a_+$ mode is $-iG\sqrt{\Gamma} \tilde{b}_{\text{th}} /(\Gamma/2 - i \omega)$ due to the optomechanical interaction. This thermal noise on the $a_+$ mode excited by the pump is in turn loaded to its degenerate mode $a_-$ through the Rayleigh backscattering channel. Therefore the effective photon occupation of the $a_-$ mode due to the pump in the cw direction becomes:
\begin{align}
	N_{th} = \frac{4 \mathcal{C} V^2}{\kappa^2 (1 + \mathcal{C})^2}n_{th}
\end{align}
where $n_{th}$ is the phonon occupation number at operation temperature, in this case at room temperature. Consequently, quantum optical effects should be observable for very large optomechanical cooperativity, in the regime $\mathcal{C} > 2 n_{th} {V^2}/{\kappa^2}$.
\vspace{24pt}

\newpage

%%%%%%%%%%%%%%%%%%%%%%%%%%%%%%%%
\subsection{Normal modes without pump backscattering (two-mode split)} 
%%%%%%%%%%%%%%%%%%%%%%%%%%%%%%%%

\begin{figure}[b!]
	\begin{adjustwidth}{-1in}{-1in}
		\makebox[\textwidth][c]{\includegraphics[width=1\textwidth]{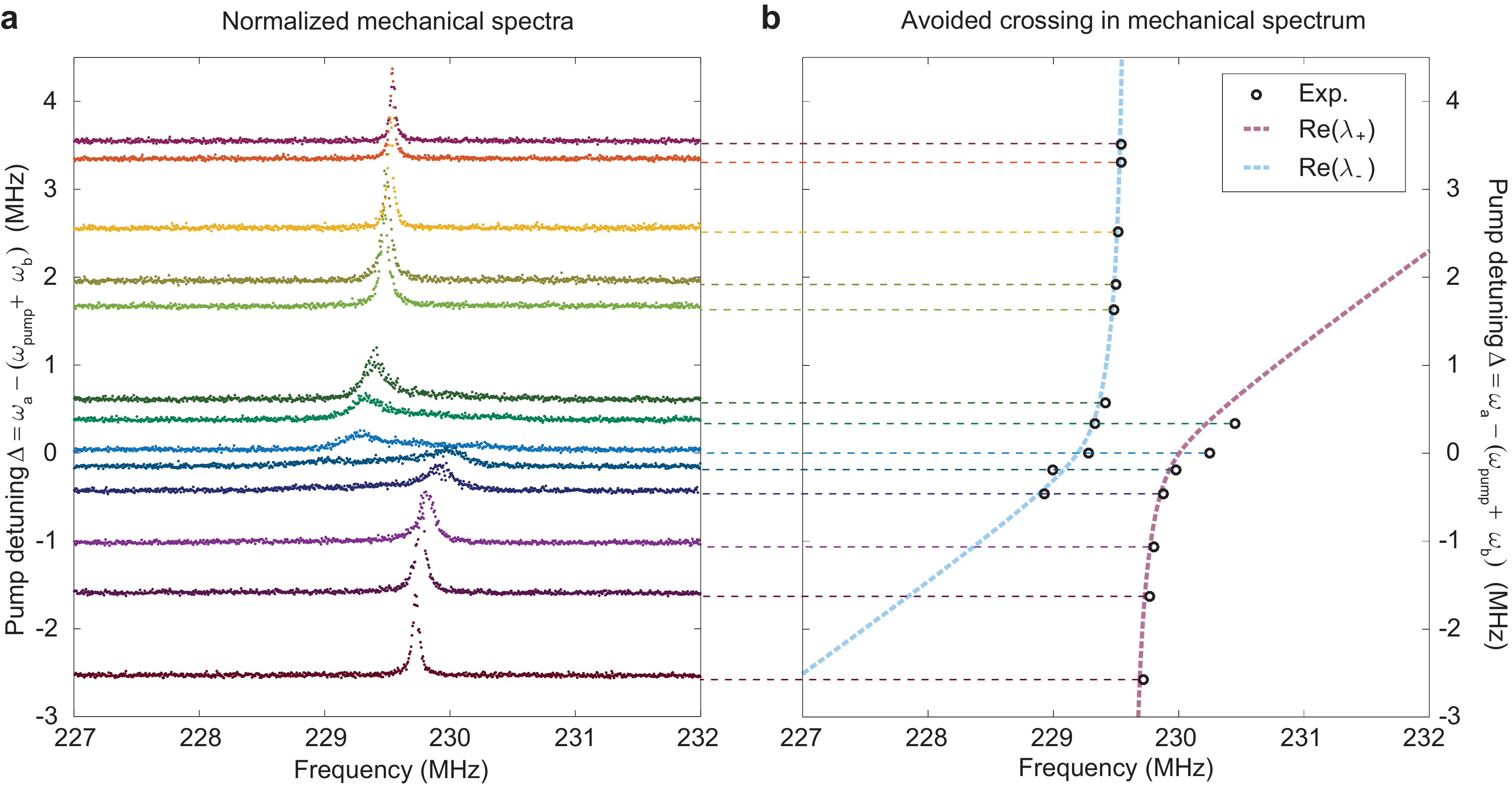}}
		\centering
		\caption{
			\textbf{Normal mode splitting with negligible pump backscattering.} 
			\textbf{(a)} We measure the mechanical spectrum through homodyne detection of the beating of the pump and scattered light. The two normal modes resulting from coupling of the $a_+$ optical and $b_+$ mechanical modes can be readily seen. This experimental measurement of mechanical spectra corresponds to Fig.~4 in the main text. 
			\textbf{(b)} The experimentally measured normal mode frequencies are compared against he theoretical curves of Eqn.~\ref{eq:eigen}. Additional details are provided in the text.	
			} 
		\label{fig:strong_coupling}
	\end{adjustwidth}
\end{figure}

The optomechanical hybridization of the optical mode $a_+$ and mechanical mode $b_+$ can be seen in either the mechanical or optical spectra. 
Optical frequency measurement of the normal modes in the strong coupling regime was presented in the main text Fig.~4.
These normal modes are also observable through the mechanical spectrum, which is presented through pump scattering measurements \cite{Bahl:2011} in Fig.~\ref{fig:strong_coupling}a.
To model this system, we chose a frame rotating with the pump laser frequency $\omega_{\text{pump}}$, i.e. we re-write $a_+ = a_+ e^{-i\omega_{\text{pump}} t }$, to obtain the mechanical frequency normal modes.
As we will show later, the pump backscattering factor $\eta$ in this experiment is small enough so as to be negligible. 
We can thus rewrite the equations of motion for the cw modes Eqns.~\ref{eq:eom} in matrix form as follows:
\begin{align}
\frac{d}{dt} 
\begin{bmatrix}
	a_+ \\ b_+
\end{bmatrix} &= -i
\begin{bmatrix}
-i \ddfrac{\kappa + 4V^2/ \kappa}{2} + (\omega_a - \omega_{\text{pump}}) & G \\
G^* &	  -i\ddfrac{\Gamma}{2} + \omega_b 
\end{bmatrix}
\begin{bmatrix}
a_+ \\ b_+
\end{bmatrix} \label{eq:eom_strong}
\end{align}
where $\kappa + 4V^2/ \kappa$ is the combined optical loss rate including the loss contribution from Rayleigh backscattering into mode $a_-$. 
The eigenvalues for the matrix in Eqn.~\ref{eq:eom_strong} are evaluated as :
\begin{align}
	\lambda_\pm = -i \frac{\kappa + 4V^2/ \kappa+ \Gamma}{4} &+ \frac{\omega_a  + \omega_b - \omega_{\text{pump}}}{2} \nonumber\\
	&\pm \frac{1}{2}\sqrt{4G^2 + \left[{\Delta}^2 + i \left(\frac{\Gamma}{2} - \frac{\kappa + 4V^2/ \kappa}{2}\right) \right]^2}\label{eq:eigen}
\end{align}
Here $\Delta = (\omega_a - \omega_b) - \omega_{\text{pump}}$ is the pump laser detuning that we also define in the main text. 
We can now obtain the normal mode frequencies by taking the real parts of the eigenvalues in Eqn.~\ref{eq:eigen}. 
Figure~\ref{fig:strong_coupling}b compares the experimental data to the theoretical prediction from this analysis. 
The dots represent the measured peak frequencies for each normal mode in Fig.~\ref{fig:strong_coupling}a. The blue and red curves show the theoretical prediction based on Eqn.~\ref{eq:eigen} for the given parameters $G=0.5$~MHz, $\kappa = 0.59$~MHz, $V = 0.3$~MHz, $\omega_b = 229.6$~MHz and $\Gamma = 39.1$~kHz, which are extracted from Fig.~4 in the main text.

The experimental results presented in Fig.~\ref{fig:strong_coupling} show only two normal-modes in the mechanical domain, corresponding to coupling of the cw $a_+$ optical mode with the cw $b_+$ mechanical mode. As we show next, this assures us that there is negligible ccw optomechanical interaction due to backscattering of the pump.

%%%%%%%%%%%%%%%%%%%%%%%%%%%%%%%%
\subsection{Normal modes including pump backscattering (four-mode split)} 
%%%%%%%%%%%%%%%%%%%%%%%%%%%%%%%%

\begin{figure}[b!]
	\begin{adjustwidth}{-1in}{-1in}
		\makebox[\textwidth][c]{\includegraphics[width=1\textwidth]{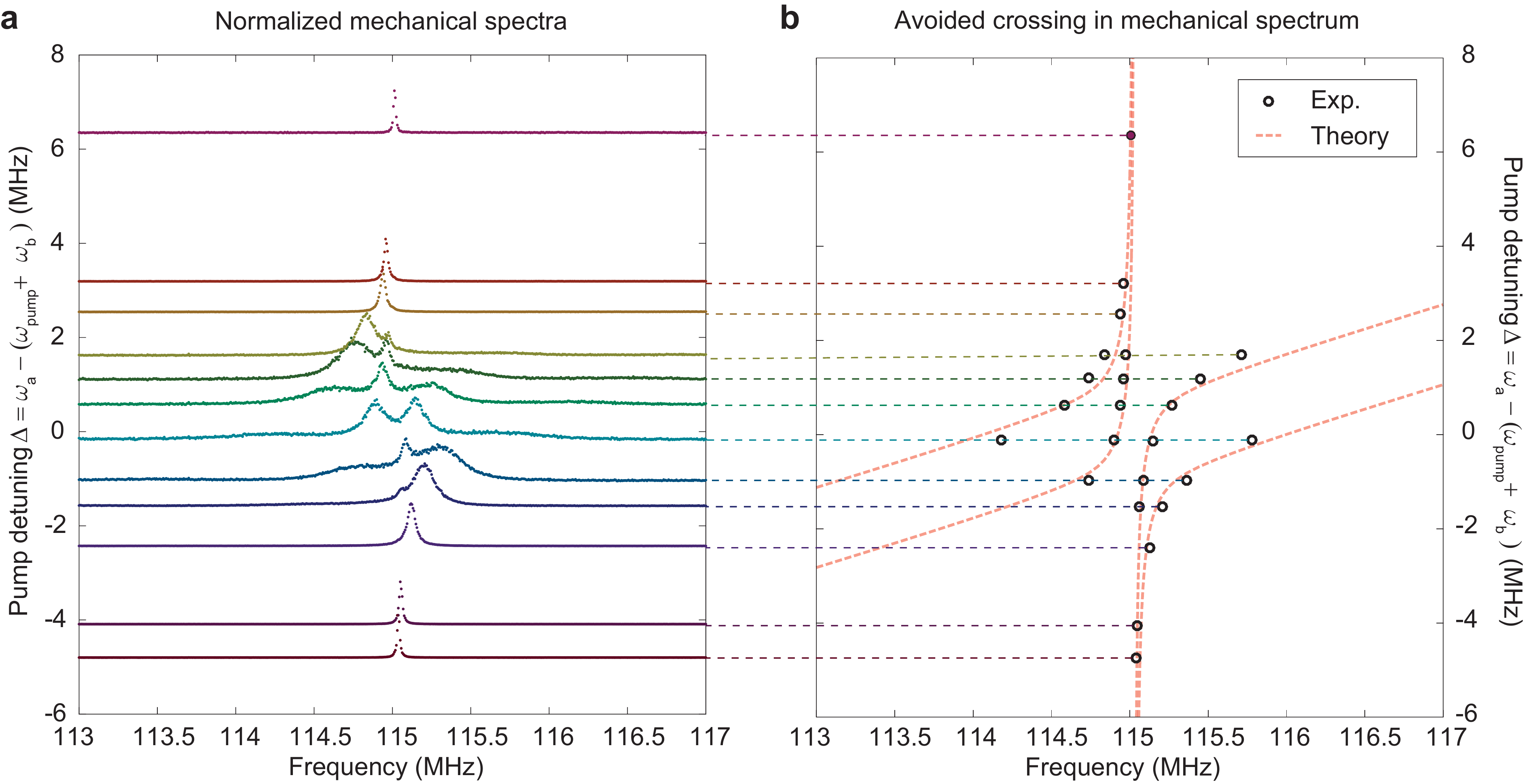}}
		\centering
		\caption{
			\textbf{Normal mode splitting with appreciable pump backscattering.} 
			\textbf{(a)} In this case, the mechanical spectra show four normal modes, produced by the optomechanical coupling of $a_\pm$ with their corresponding $b_\pm$, and through the Rayleigh scattering induced coupling of $a_\pm$.			
			\textbf{(b)} The experimentally measured normal mode frequencies are compared against the theoretical curves of Eqn.~\ref{eq:eom_strong_four}. Additional details are provided in the text.
			}
		\label{fig:strong_coupling_four}
	\end{adjustwidth}
\end{figure}

We can now also examine the corrections to the normal mode splitting that arise if pump backscattering is not negligible. 
Once again, we use the pump frequency $\omega_\text{pump}$ as reference, and rewrite the more general equations of motion (Eqn.~\ref{eq:eom}) in matrix form.
\begin{align}
\frac{d}{dt} 
	\begin{bmatrix}
		a_+ \\ b_+ \\ a_- \\ b_-
	\end{bmatrix} &= -i
	\begin{bmatrix}
		-i \ddfrac{\kappa}{2} + \Delta_a & G 			& V 	& 0 \\
		G^* 				 					& -i\ddfrac{\Gamma}{2} + \omega_b  	& 0		&0 \\
		V				 &  0 					& -i \ddfrac{\kappa}{2} + \Delta_a		& -i \eta G\\
		0 				 &  0  				& i \eta G^*						& -i\ddfrac{\Gamma}{2} + \omega_b 
	\end{bmatrix}
	\begin{bmatrix}
		a_+ \\ b_+ \\ a_- \\ b_-
	\end{bmatrix} \label{eq:eom_strong_four}
\end{align}
The coupling terms of the matrix in Eqn.~\ref{eq:eom_strong_four} imply that this system will have four normal modes. The spectra of these normal modes can be observed using the mechanical spectrum Fig.~\ref{fig:strong_coupling_four}a. 
In Fig.~\ref{fig:strong_coupling_four}b we present the analytical eigenvalue curves for the four normal modes based on Eqn.~\ref{eq:eom_strong_four}, using the parameters ($\kappa, G, V, \omega_b, \eta, \Gamma$) = (0.6~MHz, 0.4~MHz, 0.85~MHz, 115.03~MHz, 0.7, 10~kHz) that correspond to this experiment.
We have specifically avoided such cases in the experiment that we present in the main text, so as to have a clean determination of how cw pump induced time-reversal symmetry breaking affects only the $a_\pm$ modes.

\FloatBarrier

\newpage

%%%%%%%%%%%%%%%%%%%%%%%%%%%%%%%%
\section{Details on experimental measurements} 
%%%%%%%%%%%%%%%%%%%%%%%%%%%%%%%%

\subsection{Description of experimental setup}
\label{sec:Setup}

\begin{figure}[b!]
	\begin{adjustwidth}{-1in}{-1in}
		\makebox[\textwidth][c]{\includegraphics[width=0.8\textwidth]{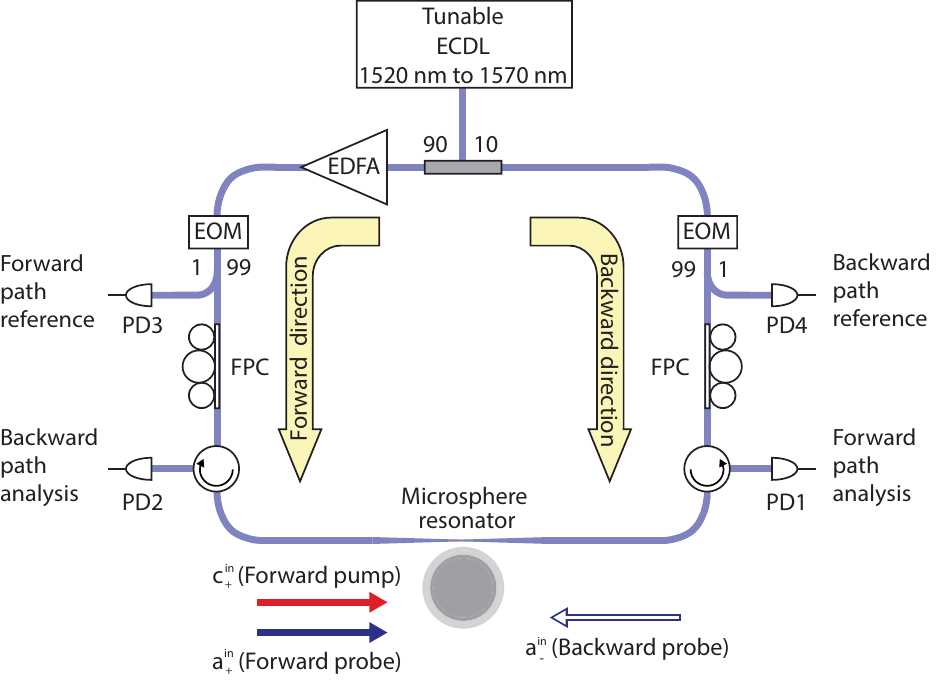}}
		\centering
		\caption{\textbf{Experimental setup details.} 
		A fiber-coupled tunable external cavity diode laser (ECDL) is sent through a 90/10 fiber splitter to produce forward and backward propagating optical signals. 
		An erbium doped fiber amplifier (EDFA) in the forward direction controls the pump laser power. 
		Electro-optic modulators (EOMs) are additionally help produce the forward and backward probes laser, while fiber polarization controllers (FPCs) are used to match polarization with the resonator modes. 
		Four photodetectors (PDs) help perform transmission and reflection measurements assisted by circulators. 
		The measured signals from the PDs are analyzed using an oscilloscope, electrical spectrum analyzer, and an electrical network analyzer.}
		\label{fig:exp_setup}
	\end{adjustwidth}
\end{figure}

The measurement setup used for our experiments is presented in Fig.~\ref{fig:exp_setup}. We employed fused-silica microsphere resonators, produced on fiber using arc discharge reflow, that are evanescently coupled to a taper fiber waveguide for probing. A 1520~nm to 1570~nm tunable external cavity diode laser was employed to drive optical pump into the waveguide. The laser source was split into the forward and backward pathways using a 90/10 splitter. 
An Erbium-doped fiber amplifier (EDFA) amplifies the pump power in the forward pathway only (i.e. for cw pumping of the resonator).
Electro-optic modulators (EOMs) are used for regulating pump power in both directions by modifying their respective dc bias, and also for producing the optical probes.
Fiber polarization controllers (FPC) are used to adjust the forward and backward probe polarizations to match the resonator modes.

During experiments, the probe sidebands generated by the EOMs are swept through the $a_\pm$ mode. The input and output fields are measured at four photodetectors; the optical signals after coupling to the resonator are collected at PD1 and PD2. Alongside, PD3 and PD4 are placed just after the EOMs to obtain 1\% of the optical signal for reference. We use two circulators for performing simultaneous measurement of the forward and backward probe transmissions and reflections.

\vspace{12pt}

%%%%%%%%%%%%%%%%%%%%%%%%%%%%%%%%%
\subsection{Calibration of optical transmission and reflection coefficients} 
%%%%%%%%%%%%%%%%%%%%%%%%%%%%%%%%%

%%%%%%%%%%%%%%%%%%%%%%%%%%%%%%%%%
\subsubsection{Transmission coefficients}
%%%%%%%%%%%%%%%%%%%%%%%%%%%%%%%%%
\label{sec:transcoeff}

Here we describe the procedure for determination of transmission coefficients of the probe signals using a network analyzer. 

\begin{figure}[b!]
	\begin{adjustwidth}{-1in}{-1in}
 	\makebox[\textwidth][c]{\includegraphics[width=1\textwidth]{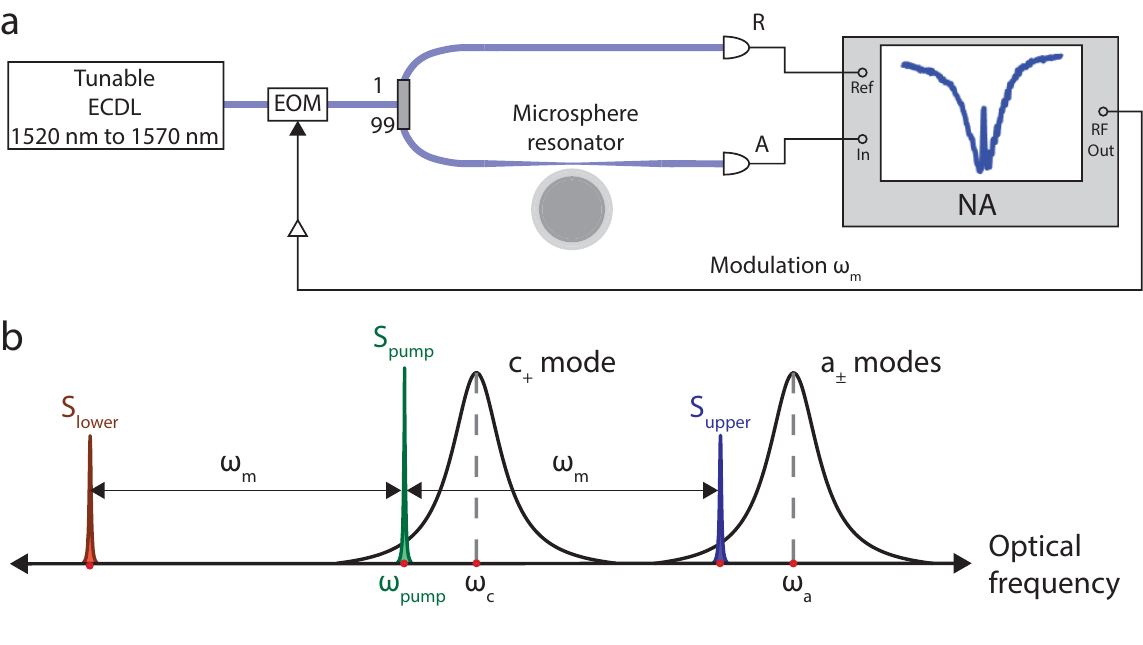}}
 	\centering
 	\caption{
		\textbf{(a)} Schematic for transmission measurements distilled from Fig.~\ref{fig:exp_setup}. 
		\textbf{(b)} The EOM produces two optical sidebands to the pump or carrier laser field ($S_\text{pump}$). While the pump is parked within the $c_+$ mode, the upper sideband is able to probe the $a_\pm$ modes. We detail the mathematics for the measurement in \S \ref{sec:transcoeff}.}
 	\label{fig:cal}
	\end{adjustwidth}
\end{figure}

For illustration purposes, the schematic of a typical transmission measurement is presented in Fig.~\ref{fig:cal}a. 
We use 1~\% of the modulated signal after the EOM as a reference at photodetector R, while 99~\% of the signal is coupled to the resonator via waveguide and its transmission is measured at photodetector A. 
Since we use an EOM to produce the probe signal (from the pump), there exist two sidebands $\omega_{\text{pump}} \pm \omega_m$ relative to the pump that propagate through the system, where $\omega_m$ is the modulation frequency received from the network analyzer. 
All optical signals and their positions in frequency space are illustrated in Fig.~\ref{fig:cal}b.
Only the upper sideband marked $S_{\text{upper}}$ measures the $a_\pm$ mode of interest, while the lower sideband marked $S_{\text{lower}}$ passes through the waveguide without interacting with the resonator. Therefore, the upper sideband is used as the optical probe.

The received optical intensities at the two photodetectors can then be expressed as:
\begin{subequations}
	\begin{align}
	&\left| E_R \, e^{-i \omega_{\text{pump}} t} \left(1+ \frac{\mathcal{B}}{2} e^{-i \omega_m t} + \frac{\mathcal{B}}{2} e^{ i \omega_m t} \right) +~\text{c.c}~\right|^2\\
	&\left| E_A \, e^{-i \omega_{\text{pump}} t} \left(t_c + t_{us} \frac{\mathcal{B}}{2} e^{-i \omega_m t} + t_{ls} \frac{\mathcal{B}}{2} e^{ i \omega_m t} \right) +~\text{c.c}~\right|^2
	\end{align}
	\label{eq:intensity}
\end{subequations}

\noindent
where $\mathcal{B}$ is the EOM intensity modulation coefficient, while $E_R$ and $E_A$ are the amplitudes of electric fields in the reference and resonator paths respectively. Here we have defined $t_c$, $t_{us}$ and $t_{ls}$ as the transmission coefficients of the optical carrier (pump), upper sideband (probe), and lower sideband signals, respectively. 
The output photocurrent is proportional to the optical intensity expressed in Eqns~(\ref{eq:intensity}). However, since the detectors have limited bandwidth in the RF domain, and the network analyzer only measures terms at frequency $\omega_m$, the only terms of interest in the output photocurrents are :
\begin{subequations}
\begin{align}
	R &= 2|E_R|^2 \mathcal{B} \cos(\omega_m t)\\
	A &=  \frac{t_c |E_A|^2 \mathcal{B}}{2} \left( e^{i \omega_m t} + t_{p} e^{-i \omega_m (t + \phi^{\prime})}\right) +~\text{c.c}
\end{align}\label{eq:AR}
\end{subequations}

\noindent
Here, we have simplified $t_c = t_c^{*}$ to set it as a reference phase, $t_{ls} = e^{  i \omega_m \phi^{\prime}}$ since the lower sideband does not interact with the resonator, and $t_{us} = t_p e^{ - i \omega_m \phi^{\prime}}$ with the new subscript indicating that it is the optical probe. 
We can now rewrite Eqn.~(\ref{eq:AR}b) as follows :
\begin{align*}
	A = \frac{t_c |E_A|^2 \mathcal{B}}{2} &\left[\left\{  (1+t_p^{\prime}) \cos(\omega_m \phi^{\prime}) +t_p^{\prime\prime} \sin(\omega_m \phi^{\prime})\right\} \cos(\omega_m t) \right.\\
	& +\left. \left\{ t_p^{\prime\prime} \cos(\omega_m \phi^{\prime}) -(1 + t_p^{\prime}) \sin(\omega_m \phi^{\prime})\right\} \sin(\omega_mt) \right]
\end{align*}
The network analyzer in the configuration of Fig.~\ref{fig:cal}a provides a complex-valued ratio of A to R. 
This result can be separated into in-phase (X) and quadrature (Y) terms as follows:
\begin{subequations}
	\begin{align}
	X = \frac{t_c M}{2} \left[ (1 + t_p^{\prime}) \cos(\omega_m \phi^{\prime}) + t_p^{\prime \prime} \sin(\omega_m \phi^{\prime})\right]\\
	Y = \frac{t_c M}{2} \left[ t_p^{\prime \prime} \cos(\omega_m \phi^{\prime}) - (1+t_p^{\prime}) \sin(\omega_m \phi^{\prime})\right]
	\end{align}
\end{subequations}
where M is a proportionality constant that includes the power split ratio (1:99), the slight difference in photodetectors' responsivities, and the difference in gain of the two optical paths. 
We can then write the calibrated probe transmission coefficient $t_m$ as follows :
\begin{align}
	t_m & = X+iY \nonumber\\
	& = \frac{t_c M}{2}\left[ (1 + t_p^{\prime} + t_p^{\prime \prime}) \cos(\omega_m \phi^{\prime}) - i(1 + t_p^{\prime} + t_p^{\prime \prime}) \sin(\omega_m \phi^{\prime})\right] \nonumber \\
	 & = \frac{t_c M}{2} (1 + t_p) e^{-i \Phi} ~.
\end{align}
Since the carrier (pump) transmission $t_c$, $M$, and the waveguide dispersion contribution $\Phi = \omega_m \phi^{\prime}$ are experimentally measurable, we can extract the true transmission coefficient $t_p$ after performing simple calibrations.

%\vspace{24pt}

%%%%%%%%%%%%%%%%%%%%%%%%%%%%%%%%%
\subsubsection{Reflection coefficients}
%%%%%%%%%%%%%%%%%%%%%%%%%%%%%%%%%

We can now similarly calibrate the optical response to obtain the reflection coefficient using a backward photodetector (via circulator) on the 99~\% branch. 
The measured optical intensity at this backward photodetector is given as : 
\begin{align}
	\left| E_A e^{-i \omega_{\text{pump}} t} \left(r_c + r_{us} \frac{\mathcal{B}}{2} e^{-i \omega_m t} + r_{ls} \frac{\mathcal{B}}{2} e^{ i \omega_m t} \right) +~\text{c.c}~\right|^2 \label{eq:Refl_A}
\end{align}
where $r_c$, $r_{us}$ and $r_{ls}$ are the reflection coefficients of carrier, upper sideband and lower sideband modes, respectively.
Once again, we set the carrier (pump) as the reference $r_c = r_c^{*}$, and since the lower sideband does not interact with the resonator we can say $r_{ls} = 0$. 
We can then rewrite the upper sideband reflection coefficient as $r_{us} = r_p e^{ - i \omega_m \phi^{\prime}}$ to indicate the probe reflection coefficient $r_p = (r_p^{\prime} + i r_p^{\prime \prime})$.
\begin{align*}
A = \frac{r_c |E_A|^2 \mathcal{B}}{2} &\left[\left\{  r_p^{\prime} \cos(\omega_m \phi^{\prime}) + r_p^{\prime\prime} \sin(\omega_m \phi^{\prime})\right\} \cos(\omega_m t) \right.\\
& +\left. \left\{ r_p^{\prime\prime} \cos(\omega_m \phi^{\prime}) -r_p^{\prime} \sin(\omega_m \phi^{\prime})\right\} \sin(\omega_mt) \right]
\end{align*}
As before, the in-phase ($X$) and quadrature ($Y$) terms from the network analyzer can be written.
\begin{subequations}
	\begin{align}
	X = \frac{r_c M}{2} \left[ r_p^{\prime} \cos(\omega_m \phi^{\prime}) + r_p^{\prime \prime} \sin(\omega_m \phi^{\prime})\right]\\
	Y = \frac{r_c M}{2} \left[ r_p^{\prime \prime} \cos(\omega_m \phi^{\prime}) - r_p^{\prime} \sin(\omega_m \phi^{\prime})\right]
	\end{align}
\end{subequations}
Once again, we can write $r_m = X + i Y$ to produce a calibrated reflection coefficient, from which the true reflection coefficient $r_p$ can be determined once $t_c$, $M$, and $\Phi = \omega_m \phi^{\prime}$ are experimentally measured.
\begin{align}
	r_m & = \frac{r_c M}{2}\left[ (r_p^{\prime} + r_p^{\prime \prime}) \cos(\omega_m \phi^{\prime}) - i(r_p^{\prime} + r_p^{\prime \prime}) \sin(\omega_m \phi^{\prime})\right] 
	 = \frac{t_c M}{2} r_p e^{-i \Phi}
\end{align}

\vspace{24pt}

\newpage
\bibliographystyle{myIEEEtran}
\bibliography{sup_ref}